\edef\TeXjobname{\jobname} % keep a copy just for safety
\edef\jobname{\detokenize{LSBD-MHONGOOSE}}
\documentclass[longauth]{aa}  
\bibpunct{(}{)}{;}{a}{}{,} % to follow the A&A style

\usepackage{wrapfig}
\usepackage{graphicx}
\usepackage{tabularx}
\usepackage{txfonts}
\def\HI{\ion{H}{I}}
\newcommand{\matHI}{{\rm H {\hskip 0.02cm \tt I}}}

\usepackage{printlen}
\usepackage{layouts}

\newcommand{\kms}{$\,$km~s$^{-1}$}

\newcommand{\ergs}{$\,$erg~s$^{-1}$}

\newcommand{\mJyb}{mJy\,beam$^{-1}$}
\newcommand{\mJy}{mJy}

\newcommand{\msun}{{${\rm M}_\odot$}}
\newcommand{\msunyr}{{${\rm M}_\odot$ yr$^{-1}$}}
\newcommand{\lsun}{{$L_\odot$}}
\newcommand{\cmsq}{cm$^{-2}$}

\newcommand{\eg}{\mbox{e.g.}}
\newcommand{\ie}{\mbox{i.e.}}

\newcommand{\mhon}{{MHONGOOSE}}

\newcommand{\lsb}{\mbox LSB-D}

\newcommand{\nfi}{\mbox NGC~1566}

\newcommand{\meer}{{MeerKAT}}

\newcommand{\cara}{{\tt CARACal}}

\newcommand{\vsys}{{$v_{\rm sys}$}\,}

\newcommand{\nhi}{{$N_{\rm H {\hskip 0.02cm \tt I}}$}}
\newcommand{\mhi}{{$M_{\rm H {\hskip 0.02cm \tt I}}$}}

\begin{document}

\title{MHONGOOSE discovery of a gas-rich low-surface brightness galaxy in the Dorado Group}   
	\titlerunning{\HI\ rich LSB in the Dorado group}
	\authorrunning{F. M. Maccagni, et al.}
 
       \author{F.~M. Maccagni\inst{1,2}, 
       W.~J.~G. de Blok\inst{1,3,4}, 
       P.~E.~Mancera Pi\~na\inst{5}, 
       R.~Ragusa\inst{6}, 
       E.~Iodice\inst{6},
       M.~Spavone\inst{6},
       S.~McGaugh\inst{7},
       K.~A.~Oman\inst{8,9,10},
       T.~A.~Oosterloo\inst{1,4},
       B.~S.~Koribalski\inst{11,12},
       M.~Kim\inst{13},
       E.~A.~K. Adams\inst{1,4},
       P.~Amram\inst{14},
       A.~Bosma\inst{14},
       F.~Bigiel\inst{15},
       E.~Brinks\inst{16},
       L.~Chemin\inst{17},
       F.~Combes\inst{18,19},
       B.~Gibson\inst{20},
       J.~Healy\inst{1},
       B.~W.~Holwerda\inst{21}, 
       G.~I.~G.~J\'ozsa\inst{22,23},
       P.~Kamphuis\inst{24},
       D.~Kleiner\inst{1,2},
       S.~Kurapati\inst{3},
       A.~Marasco\inst{25},
       K.~Spekkens\inst{26,27},
       S.~Veronese\inst{1,4},
       F.~Walter\inst{28},
       N.~Zabel\inst{3},
       A.~Zijlstra\inst{29}}
         \institute{$^1$ ASTRON - Netherlands Institute for Radio Astronomy, Oude Hoogeveensedijk 4, 7991 PD, Dwingeloo, The Netherlands \\
        $^2$ INAF -- Osservatorio Astronomico di Cagliari, via della Scienza 5, 09047, Selargius (CA), Italy\\
        $^3$ Dept. of Astronomy, Univ. of Cape Town, Private Bag X3, Rondebosch 7701, South Africa\\
	    $^4$ Kapteyn Astronomical Institute, University of Groningen, PO Box 800, 9700 AV Groningen, The Netherlands\\
        $^5$ Leiden Observatory, Leiden University, P.O. Box 9513, 2300 RA Leiden, The Netherlands\\
        $^6$ INAF -- Osservatorio Astronomico di Capodimonte, via Moiariello 16, Napoli 80131, Italy\\
        $^7$ Department of Astronomy, Case Western Reserve University, Cleveland, OH 44106, USA\\        
        $^8$ Institute for Computational Cosmology, Durham University, South Road, Durham DH1 3LE, United Kingdom\\
        $^9$ Centre for Extragalactic Astronomy, Durham University, South Road, Durham DH1 3LE, United Kingdom\\
        $^{10}$ Department of Physics, Durham University, South Road, Durham DH1 3LE, United Kingdom\\
        $^{11}$ Australia Telescope National Facility, CSIRO Astronomy and Space Science, P.O. Box 76, Epping, NSW 1710, Australia \\
        $^{12}$ Western Sydney University, Locked Bag 1797, Penrith South, NSW 1797, Australia\\
        $^{13}$ Department of Astronomy and Space Science, Sejong University, Seoul 05006, Korea\\
        $^{14}$ Aix Marseille Univ, CNRS, CNES, LAM, Marseille, France\\
        $^{15}$ Argelander-Institut f\"ur Astronomie, Auf dem H\"ugel 71, 53121, Bonn, Germany\\
        $^{16}$ Centre for Astrophysics Research, University of Hertfordshire, College Lane, Hatfield, AL10 9AB, UK\\   
        $^{17}$ Instituto de Astrof\'isica, Departamento de Ciencias F\'isicas, Universidad Andr\'es Bello, Fernandez Concha 700, Las Condes, Santiago, Chile\\
        $^{18}$ LERMA, Observatoire de Paris, PSL research Université, CNRS, Sorbonne Universit\'e, 75104, Paris, France, \\
        $^{19}$ Coll\'ege de France, 11 Place Marcelin Berthelot, 75005, Paris, France,\\
        $^{20}$ E.A. Milne Centre for Astrophysics, University of Hull, Hull, HU6 7RX, United Kingdom\\
        $^{21}$ University of Louisville, Department of Physics and Astronomy, 102 Natural Science Building, 40292 KY Louisville, USA,\\
        $^{22}$ Max-Planck-Institut f\"ur Radioastronomie, Auf dem H\"ugel 69, 53121, Bonn, Germany\\
        $^{23}$ Department of Physics and Electronics, Rhodes University, PO Box 94, Makhanda, 6140, South Africa\\
        $^{24}$ Ruhr University Bochum, Faculty of Physics and Astronomy, Astronomical Institute (AIRUB), 44780 Bochum, Germany,\\
        $^{25}$ INAF -- Padova Astronomical Observatory, Vicolo dell’Osservatorio 5, I-35122 Padova, Italy\\
        $^{26}$ Department of Physics and Space Science, Royal Military College of Canada P.O. Box 17000, Station Forces Kingston, ON K7K 7B4, Canada\\
        $^{27}$ Department of Physics, Engineering Physics and Astronomy, Queen’s University, Kingston, ON K7L 3N6, Canada\\
		$^{28}$ Max-Planck-Institut f\"{u}r Astronomie, K\"{o}nigstuhl 17, D-69117, Heidelberg, Germany,\\
        $^{29}$ Department of Physics and Astronomy, The University of Manchester, Manchester M13 9PL, UK\\ 
        \\
   \email{filippo.maccagni@inaf.it}
             }

   \date{Received 01 February 2024 / Accepted 25 May 2024}

  \abstract
   {We present the discovery of a low-mass gas-rich low-surface brightness galaxy in the Dorado Group, at a distance of 17.7 Mpc. Combining deep MeerKAT 21-cm observations from the MeerKAT \HI\ Observations of Nearby Galactic Objects: Observing Southern Emitters (MHONGOOSE) survey with deep photometric images from the VST Early-type Galaxy Survey (VEGAS) we find a stellar and neutral atomic hydrogen (\HI) gas mass of $M_\star = 2.23\times10^6$~\msun\ and \mhi$=1.68\times10^6$~\msun, respectively. This low-surface brightness galaxy is the lowest mass \HI\ detection found in a group beyond the Local Universe ($D\gtrsim 10$ Mpc). The dwarf galaxy has the typical overall properties of gas-rich low surface brightness galaxies in the Local group, but with some striking differences. Namely, the MHONGOOSE observations reveal a very low column density ($\sim 10^{18-19}$~\cmsq) \HI\ disk with asymmetrical morphology possibly supported by rotation and higher velocity dispersion in the centre. There, deep optical photometry and UV-observations suggest a recent enhancement of the star formation. Found at galactocentric distances where in the Local Group dwarf galaxies are depleted of cold gas (at $390$ projected-kpc distance from the group centre), this galaxy is likely on its first orbit within the Dorado group. We discuss the possible environmental effects that may have caused the formation of the \HI\ disk and the enhancement of star formation, highlighting the short-lived phase (a few hundreds of Myr) of the gaseous disk, before either SF or hydrodynamical forces will deplete the gas of the galaxy.}


   \maketitle

\section{Introduction}
\label{sec:intro}

In the hierarchical model of galaxy evolution, dwarf galaxies ($M_\star\lesssim 10^9$~\msun) are the most abundant in the Universe and form the building blocks of more massive galaxies, such as our own Milky Way (MW). Dwarf galaxies typically have dynamical masses more than four times larger than their baryonic mass~\citep[\eg][]{Das:2020}, which makes them an ideal laboratory to test cosmological evolution and models dark matter (DM), which counts to up to $\sim 27\%$ of the total energy density of the Universe~\citep[\eg][]{PlanckCollaboration:2020}. Nevertheless, little is known about the formation and evolution of these low-mass galactic systems. Dwarf galaxies can be strongly and rapidly affected by their environment. Tidal interactions, for example, are known to generate a specific class of dwarf galaxies (\ie\ tidal dwarfs) which in their formation are depleted of DM~\citep[\eg][]{Duc:1998,Read:2017,venhola:2017,Buck:2019,Gray:2023}. Conversely, hydro-dynamical interactions with the inter-galactic medium (IGM) can  strip and deplete the gaseous disks of these galaxies~\citep[\eg][]{McConnachie:2007,Westmeier:2011,Jozsa:2021,Yang:2022}. Dwarfs then help us build a complete picture on how the environment affects the availability of their gas reservoir, ultimately regulating their star formation history. They are fundamental to understand the dynamical processes driving the assembly of the stellar and gaseous disks of galaxies.

The new generation of wide-field optical imaging surveys, such as the Dark Energy Survey~\citep[DES,~\eg][]{Dey:2019,Martinez-Delgado:2023} or the VST Early-type Galaxy Survey~\citep[VEGAS,~\eg][]{capaccioli:2015,iodice:2020} have enabled the discovery and study of the stellar component of hundreds of dwarf galaxies down to very low surface brightnesses, i.e. $\sim 27-29$ mag/arcsec$^2$ in the g-band~\citep{forbes:2020,Iodice:2021}. However, the analysis of the gaseous component, which is the best tracer for the kinematics and consequently the DM halo and environment, has been much more difficult. A galaxy at the low-mass end of the population (M$_\star \lesssim 10^7$~\msun), with a gas fraction of $50\%$, would require a sensitivity to neutral atomic gas masses of M$_{\rm gas} \lesssim 5\times 10^6$~\msun. Hence, a search for the \HI\ in these sources has been limited so far only to the nearby Universe ($\lesssim 10$ Mpc) and by the sensitivities of \HI\ surveys such as, for example, the Westerbork observations of Neutral Hydrogen in Irregular and SPiral galaxies~\citep[WHISP, ][]{Swaters:2002}, the Very Large Array survey of Advanced Camera for Surveys Nearby Galaxy Survey Treasury galaxies~\citep[VLA-ANGST, ][]{Ott:2012}, the `(almost) dark galaxies in the ALFALFA survey'~\citep[][]{Leisman:2017}, The Local Volume \HI\ Survey~\citep[LVHIS][]{Koribalski:2018}, Apertif~\citep{Adams:2022} and the Wide-field ASKAP L-band Legacy All-sky Blind Survey~\citep[WALLABY][]{westmeier:2022}. Only in a handful of dwarf galaxies have resolved studies of the bulk of their \HI\ kinematics have been possible, such as, for example, in the sample of The \HI\ Nearby Galaxy Survey of the Local Irregulars That Trace Luminosity Extremes~\citep[LITTLE THINGS][]{Hunter:2012,oh:2015,Iorio:2017} the dwarf galaxies of LVHIS and VLA-ANGST~\citep[][]{Mancera2021a}, or in, for example, Leo T~\citep[][]{Ryan-Weber:2008,adams:2018} and Leo-P~\citep[][]{bernstein-cooper:2014}. 

Recently, the combination of deep 21-cm \HI\ observations (\nhi$\lesssim10^{19}$\cmsq) with deep optical photometry ($\mu_r\gtrsim27$ mag/arcsec$^2$) allowed us to discover several new gas-rich and low-mass low-surface brightness objects beyond the Local Group (LG). The MeerKAT Fornax Survey~\citep[][]{Serra:2023}, for example, has investigated the \HI\ counterparts of all dwarf galaxies in the Fornax cluster ($D\sim 20$ Mpc) down to an \HI\ mass limit of \mhi$\sim5\times10^5$~\msun~\citep[][]{Kleiner:2023}, that were discovered and catalogued by the optical Fornax Deep Survey down to a surface brightness limit of $\sim 26$ mag/arcsec$^2$ in the r-band~\citep[][]{venhola:2017,Venhola:2021}. 

In clusters, the environment strongly affects the \HI\ content of dwarf galaxies. In Virgo, for example, the \HI\ disks of dwarfs are stripped by ram-pressure (RP) by the IGM even in the low-density outskirts of the cluster~\citep[][]{Boselli:2008a,Boselli:2008b}. In the less dense medium of Fornax, tidal and hydrodynamical interactions efficiently deplete the \HI\ in dwarf galaxies in between a few tens and a few hundreds of Myr~\citep[][]{Kleiner:2023}. The study of low-mass objects in groups and in the field is so far more incomplete. The main limitation being the difficulty in reaching very deep optical and \HI\ surface brightness limits and in covering, at the same time, a wide enough field of view to be able to discover these sources, which are often found hundreds of kilo-parsecs away from their host galaxies. The MeerKAT \HI\ Observations of Nearby Galactic Objects: Observing Southern Emitters survey~\citep[MHONGOOSE,][]{deBlok:2016,deBlok:2020,deBlok:2024} is now providing us with the unique opportunity to discover low \HI\ mass sources (\mhi$\sim 10^6$~\msun) within $1\times 1$ deg$^2$ of 30 nearby star forming galaxies (median distance $D=10.3$ Mpc). Because of the sensitivity to diffuse \HI\ emission over the full field of view guaranteed by \meer, MHONGOOSE investigates the gaseous environment of these galaxies out to almost their virial radius. The coincidence of the \HI\ gas detected in the environment with the optical counterpart detected by sensitive optical observations ($\mu_r\gtrsim27$ mag/arcsec$^2$) allows us to construct a complete census of the low-mass gas-rich galaxy population around the target galaxies. The MHONGOOSE galaxies are mostly isolated or live in nearby sparse groups and span an \HI\ mass range between $10^6$ and $10^{11}$~\msun. The first data release of the shallow observations of the first $10\%$ of the MHONGOOSE data already mapped tens of companion satellite galaxies~\citep[][]{deBlok:2024}.

In this paper, we present the discovery of the lowest \HI\ mass system (\mhi$=1.68\times10^6$~\msun) with optical counterpart detected in the Dorado group~\citep[$D = 17.69$~Mpc,~\eg
][]{rampazzo:2020} by the MHONGOOSE deep observations of star-forming spiral galaxy \nfi. This group was observed in the neutral hydrogen by previous surveys ~\citep[\eg][]{kilborn:2005,elagali:2019}, but never with the depth and resolution provided by MHONGOOSE, which lead to the discovery of this and other \HI\ counterparts of low-mass galaxies in the group. The newly discovered \HI\ source ({\mbox MKT 042326-551621}) is the cold gas counterpart of the low-surface brightness galaxy (LSBG) catalogued, for example, in~\citet{Tanoglidis:2021} with source number 19696. The authors associated this source to the density peak of their distribution of LSBGs in the surroundings of the Fornax cluster, at 19 Mpc. For the very first time,  because of the HI systemic velocity of this source, we can determine that \lsb\ lies in the outskirts of the Dorado group. Given its location in Dorado, and its low-surface brightness nature we name this galaxy LSBG-Dorado-A, and for brevity, we hereafter refer to it as \lsb. 

Deep optical observations from VEGAS of the Dorado group allow us to study the optical counterpart  of the \HI\ detection. This galaxy is the least-massive low-surface brightness galaxy ($M_\star\sim2.45\times 10^6$~\msun) with highest gas fraction ($f_{\rm gas}= M_{\rm gas}/(M_{\rm gas}+M_\star)=0.48$) detected beyond the Local Group. Thanks to the high sensitivity and resolution of both MeerKAT and VST observations we are able to resolve and study in detail the \HI\ gas of such a small object in Dorado and relate its morphology and kinematics to its star formation history and evolution within the group's environment. 

This paper is structured as follows: in Sect.~\ref{sec:obs}, we present the \meer\ \HI\ and optical VST observations and show the morphology and kinematics of the \HI\ in \lsb\ and its optical and star formation properties. In Sect.~\ref{sec:disc} we discuss the location of this dwarf galaxy within the Dorado group, its baryon content and we present possible scenarios on the evolution of its \HI\ disk and SF. Section~\ref{sec:conc} summarizes the results and examines the possible studies of low-mass sources that are being opened by these novel deep \HI\ and optical observations.

\section{Observations and results}
\label{sec:obs}

\subsection{\mhon\ neutral hydrogen observations}
\label{sec:obsHI}

The 21-cm \meer~\citep{camilo:2018,mauch:2020} observations presented in this paper were taken as part of the MHONGOOSE survey targeting the massive lopsided spiral galaxy \nfi~\citep[\mhi=$1.1\times 10^{10}$~\msun, $D_{\rm NGC1566}=17.6$ Mpc][]{elagali:2019,anand:2021}. \nfi\ was observed for 55 hours in ten separate 5.5 hour tracks reaching \HI\ column density sensitivities of \nhi($3\sigma$)$\sim6\times 10^{18}$~\cmsq\ at 30\arcsec\ resolution, over 16~\kms. 

 \begin{figure*}[tbh]
	\centering
	\includegraphics[width=\textwidth]{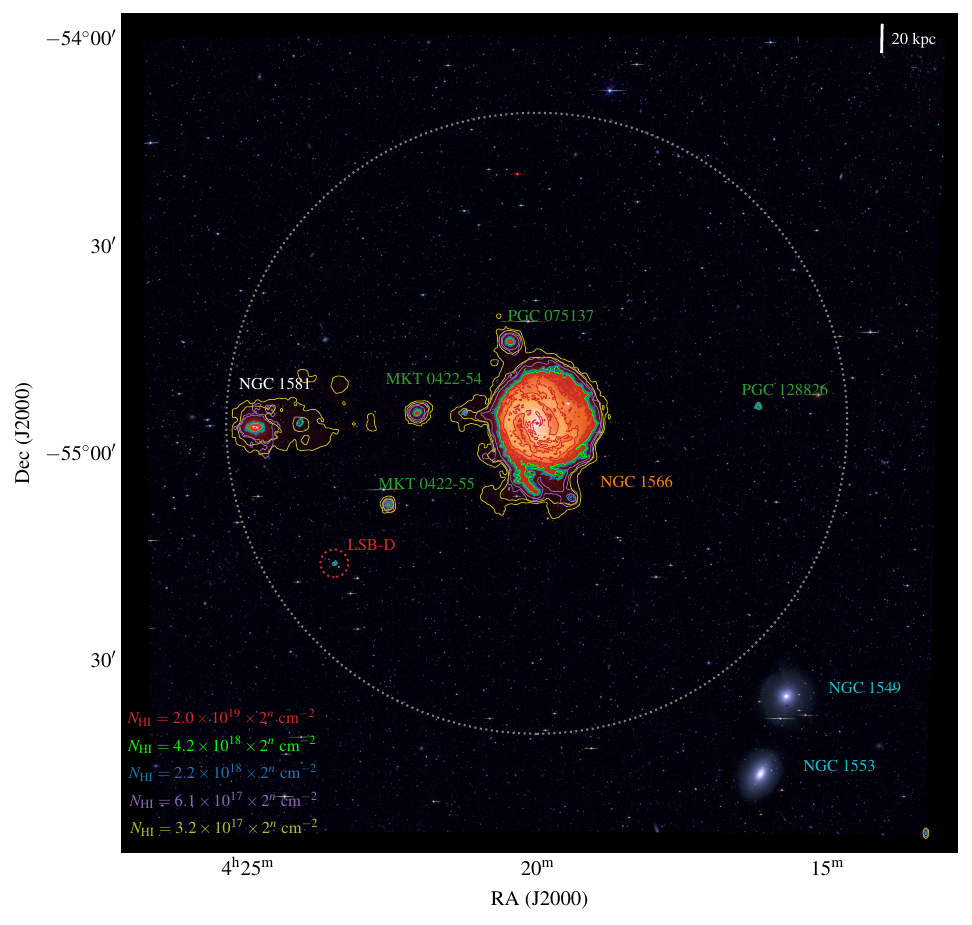}	
	\caption{Primary beam corrected flux-density \HI\ emission detected by \meer\ within a $1.5\times 1.5$ deg$^2$ field centred on \nfi\ (the imaged f.o.v. is marked by the dashed grey circle), overlaid on the DECaLS optical image in the g,r,z filters. The \HI\ emission is a composite of flux-density maps derived from the multi-resolution datacubes. The resolutions shown by coloured contours are $12$\arcsec$\times10$\arcsec\ (red), $25$\arcsec$\times18$\arcsec (green), $32$\arcsec$\times23$\arcsec\ (blue), $65$\arcsec$\times64$\arcsec\ (purple), $94$\arcsec$\times92$\arcsec\ (yellow-green). For resolutions between $\sim 90$\arcsec\ and $25$\arcsec\ two contours are shown, \ie, \nhi$=3\sigma\times2^n$ with $n = 0, 2$. The $12$\arcsec$\times10$\arcsec\ resolution has further increasing contours ($n=0,2,4$...). The PSF of the multi-resolution maps are shown in the bottom-right corner with the same colour coding. Details on the \HI\ properties of the sources in the field are given in Table~\ref{tab:masses}. }
	\label{fig:fullFieldCont}
\end{figure*}

A complete description of the data reduction of the \mhon\ observations is given in the survey paper~\citep[][]{deBlok:2024}. Here, we summarise the main steps taken to produce the \HI\ datacubes\footnote{the datacubes span approximately $720$ channels  (or a $1000$~\kms\ velocity range) centred on the target galaxy.} that led to the detection of the gas rich low-surface brightness galaxy \lsb. We reduce the data using \cara~\citep{jozsa:2020}, a containerised pipeline written in {\tt Python3}. \cara\ is set up in a modular fashion allowing us to use tasks from different radio astronomical packages for each step of the data reduction. The data-reduction steps are performed on a single node of the MeerGas cluster (128 cores,  1 Tb RAM). All ten tracks are reduced following the same strategy and using the same set of parameters, except for a few minor exceptions. 

The data reduction procedure is divided in three parts. The first part is performed on the single 5.5 hours tracks and consists of cross-calibration, RFI-flagging, self-calibration and continuum subtraction. The second part consists of combining all ten observations together in the {\em uv-}space, followed by the joint deconvolution and imaging to produce \HI\ datacubes at multiple resolutions with a 1.4~\kms\ channel width, without further continuum subtraction. In the third part, we create the moment maps (\ie\ \HI\ flux density, velocity field and \HI\ line-width map) of the imaged field of view ($1.5\times1.5$ deg$^2$) using the automated source finder {\tt SoFiA-2}~\citep[][]{westmeier:2021}.

We produce the final datacubes with six weighting and tapering combinations, that span resolutions from $90$\arcsec\ to $7$\arcsec, with $3\sigma$ sensitivity limits ranging from $5\times10^{17}$~\cmsq\ to $5\times10^{20}$~\cmsq\ over $16$~\kms, respectively (further details are given in Appendix~\ref{app:a1}). We use {\tt SoFiA-2} to search for \HI\ detections with a smooth \&\ clip algorithm, with Gaussian kernels $1$, $2$ and $3$ times the spatial resolution of the cubes. The detection threshold is set to $3.5$ times the noise (evaluated in 3D over a spectral window of 7 channels). All \HI\ sources detected in the field of view have a detection reliability of $85$\%, based on the distribution of positive and negative noise peaks in the datacubes~\citep[see][for details on the reliability estimate]{serra:2015,westmeier:2021}. We verify that {\tt SoFiA-2} identified all genuine \HI\ emission by also visually inspecting the datacubes in virtual reality~\citep[iDaVIE]{jarrett:2021}.

The \HI\ flux density distribution of \nfi\ and of the galaxies detected by \meer\ at resolutions between $90$\arcsec to $7$\arcsec\ is shown in Fig.~\ref{fig:fullFieldCont} over a combination of the $g,r,z$ optical observations of the Dorado Group from the Dark Energy Camera Legacy Survey (dr10-DECaLS). The doublet of early-type galaxies \mbox{NGC 1549} and {NGC 1553} in the centre of the group is shown in the south-west, beyond the imaged \meer\ field of view (dashed circle). New \HI\ detections are marked in green. The first contour of each colour (from yellow-green to red) shows the S/N$=3$ \HI\ column density limit of datacubes at resolutions between $90$\arcsec and $12$\arcsec. \meer\ has a primary beam diameter of 1 degree (at this diameter the sensitivity of the array drops to $50$\%). This enables us to study not only \nfi\ and its interactions with its nearby satellites, but also to investigate in detail its environment out to $\sim 360$ projected-kpc, which corresponds approximately to $75\%$ of its radius at $500$ times the critical density of the Universe at the current epoch~\citep[$r_{500}=0.47$ Mpc,][]{kilborn:2005}. 

The \HI\ MHONGOOSE observations of \nfi\ are the deepest and highest resolution 21-cm observations available so far of this galaxy (a summary of the sensitivities is shown in Table~\ref{tab:cubes}). They improve our characterization of the low-column densities features of its \HI\ disk of at least an order of magnitude~\citep[][]{elagali:2019} and allow us to detect \HI\ associated with several satellites, as well as diffuse \HI\ tails tracing past interactions (as in \mbox{NGC 1581}, in the east of the field of view, Fig.~\ref{fig:fullFieldCont}).  A full description of the cold gas distribution and kinematics of \nfi\ and the other detections will be given in a following paper, a summary of their \HI\ properties is shown in Table~\ref{tab:masses}. 

Among the new sources, to the south-east of \nfi, at 186 kpc in projection at coordinates RA $=04^{\rm h}$23$^{\rm m}$26$^{\rm s}$, Dec$= -55^{\rm d}$16$^{\rm m}$21$^{\rm s}$, lies the lowest \HI\ detection associated with a galaxy made by MHONGOOSE so far, the aforementioned \lsb\ (circled in red in Fig.~\ref{fig:fullFieldCont}). Deep optical observations allow us to identify that the \HI\ gas is co-located with a low-surface brightness galaxy. The \HI\ systemic velocity of \lsb\ (\vsys$=1214$~\kms) is consistent with the average recession velocity of the Dorado group ($v_{\rm hel}=1269$~\kms) and the source seems to be located at its outskirts. Dorado is a loose group with centre coincident with the S0 galaxy \mbox{NGC 1553} (visible in the south-west in Fig.~\ref{fig:fullFieldCont}), which forms a pair with the brightest group galaxy (BGG) \mbox{NGC 1549}~\citep[][]{Firth:2006,kourkchi:2017}. \lsb\ is located to the north-east of Dorado's centre ($68$\arcmin\ away, see Fig.~\ref{fig:fullFieldCont}), whereas \nfi\ and its companions are at a distance of $35$\arcmin. A summary of the main properties of Dorado is shown in Table~\ref{tab:Dorado}. 

\begin{table}[tbh]
        \caption{Properties of the Dorado group}
        \centering
        \label{tab:Dorado}
        \begin{tabularx}{\columnwidth}{X l}  
                \hline\hline                   
                	Parameter   & Value \\
                	\hline                                   
                	Centre RA [J2000] &  $04^{\rm h}16^{\rm m}10^{\rm s}$ \\
					Centre Dec [J2000] &  $-55^{\rm d} 46^{\rm m} 48^{\rm s}$\\
					Group Members & 31 \\
                    Morphological classification$^\dag$ & 16 ETG $/$ 10 LTG $/$ 5 Unlc.\\
                    $\langle v_{\rm hel}\rangle$ [\kms] & $1231$ \\
                    $D$ [Mpc] & $17.69$ \\
                    $r_{\rm vir}$ [kpc] & $654$ \\ 
                    $M_{\rm vir}$ [\msun] & $3.50\times 10^{13}$  \\ 
                    $\sigma_{\rm group}$ [\kms] & $282$ \\     
                	\hline                           
        \end{tabularx} 
         \tablefoot{The listed properties are taken from \citet{kourkchi:2017}, where formulae and definitions to the parameters are provided. The centre of Dorado coincides with S0 galaxy~\mbox{NGC 1553}, which has as close-by companion the BGG \mbox{NGC 1549}. The reported group distance and heliocentric velocity are based on the unweighted averaging of the radial velocities of their members.$^\dag$ Morphological classification of group members divided in early-type galaxies (ETG), late-type galaxies (LTG) and unclassified (Uncl.)~\citep{rampazzo:2020}.}
\end{table}

The \HI\ disk of \lsb\ is detected at resolutions between $10$ and $40$ arcseconds with peak signal-to-noise $S_{\rm peak}/N\gtrsim 5$, while the gas becomes too beam-diluted to be detected at the lowest resolutions and below the detection limit at the highest resolution of our datacubes ($7$\arcsec). This highlights the compact but diffuse nature of the \HI\ disk, see Sect.~\ref{sec:dm} for further details. In what follows, we analyse the results only from the $25$\arcsec $\times 18$\arcsec\ resolution datacube, with $1.4$~\kms channel-width generated with no tapering and a robustness parameter of $1.0$,~\citep[][see Table~\ref{tab:cubes}]{Briggs:1999}. At this resolution, we fully recover the total extent of the \HI, spatially resolving it with 3 beam elements, while at higher resolutions we resolve out part of the disk. The noise in the primary beam corrected cube close to \lsb\ is $0.265$~\mJyb. In Appendix~\ref{app:a2} we show the channel maps of the \HI\ emission of \lsb.

\subsection{\HI\ properties of \lsb}
\label{sec:HI}

\begin{figure}[tbh]
	\centering
	\includegraphics[width=\columnwidth]{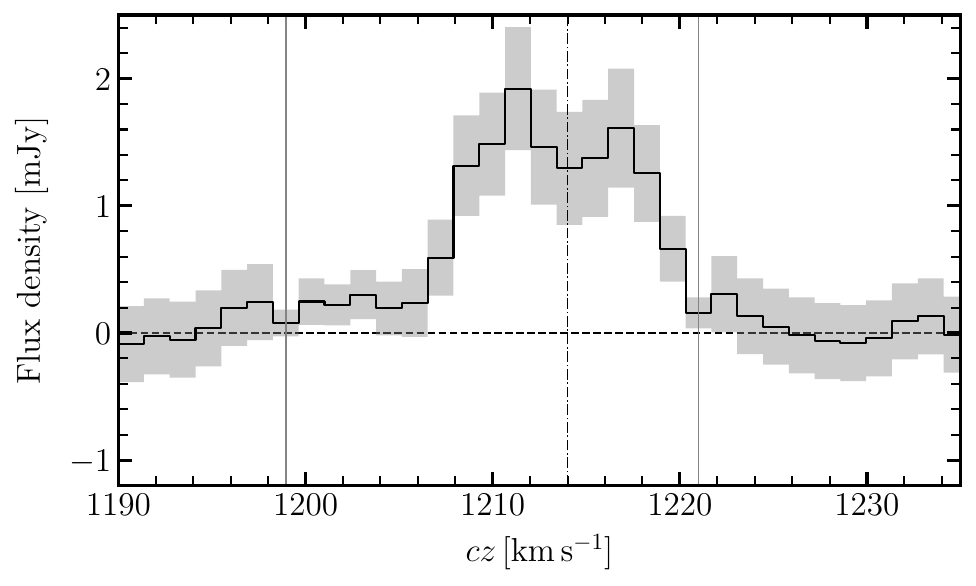}	
	\caption{Integrated spectrum of \lsb, with its noise per channel (grey shaded region). The vertical dashed line shows the systemic velocity of the source ($v^{\rm radio}_{\rm sys}=1214$\kms), while the grey vertical lines show its full velocity range ($w_{\rm FWZI}=24$\kms).}
	\label{fig:integratedSpectrum}%
\end{figure}

The integrated \HI\ spectrum  of \lsb\ is shown in Fig.~\ref{fig:integratedSpectrum}. The \HI\ line spans over 17 channels (1.4~\kms\ wide) with a full-width at zero intensity (FWZI) of $w_{\rm FWZI}=24$~\kms. The profile is double-peaked suggesting the gas may be rotating within the galaxy and an extended tail is present at blueshifted velocities. The total flux density of the \HI\ line is $S_{1.4\,{\rm GHz}}\Delta v = 22.9$~mJy~\kms. The \HI\ properties of \lsb\ are summarized in Table~\ref{tab:properties}. 

In a group, the projected line of sight velocity we measure from the \HI\ line is the composition of the systemic velocity of the source and its relative projected motion within the group. The exact trajectory of \lsb\ is unknown, but since its systemic velocity ($v_{\rm sys,\, LSB-D} =1214$~\kms) is similar to that of the group centre ($v_{\rm sys,\,NGC 1553} =1201$~\kms), and of the close-by BGG \mbox{NGC 1549} ($v_{\rm sys,\, NGC 1549} =1202$~\kms), it is reasonable to approximate its distance by that of the group centre ($D=17.69$ Mpc)\footnote{At this distance, the image scale is $85$ parsec/arcsec assuming a $\Lambda$CDM cosmology, with Hubble constant $H_0 = 70$~\kms Mpc$^{-1}$ and $\Omega_\Lambda = 0.7$ and $\Omega_M = 0.3$}, which has been estimated by averaging the radial velocities of their group members~\citep{kourkchi:2017,rampazzo:2020}. At this distance \lsb\ lies in projection $351$ kpc away from \mbox{NGC 1553} and $181$ kpc and $116$ kpc from \nfi\ and \mbox{NGC 1581}, respectively ($v_{\rm sys,\, NGC1566}=1496$~\kms, $v_{\rm sys,\, NGC1581}=1609$~\kms). Further information on the location of \lsb\ within the group is given in Sect.~\ref{sec:location}.

Following standard equations~\citep[\eg][]{meyer:2017}, the total flux of \lsb\ corresponds to an \HI\ mass of \mhi$=1.68\times10^6$~\msun. So far, this is the lowest \HI\ mass that has been associated with a low-surface brightness galaxy beyond the Local Group. For comparison, the typical dwarf population in the LG has associated \HI\ masses two orders of magnitudes higher than in \lsb, with only a few sources with \mhi$\sim 10^6$~\msun~\citep[\eg,][]{Bouchard:2006,Grcevich:2009,Hunter:2012,Spekkens:2014,Iorio:2017}. 

\begin{figure*}
	\centering
	\includegraphics[width=\textwidth]{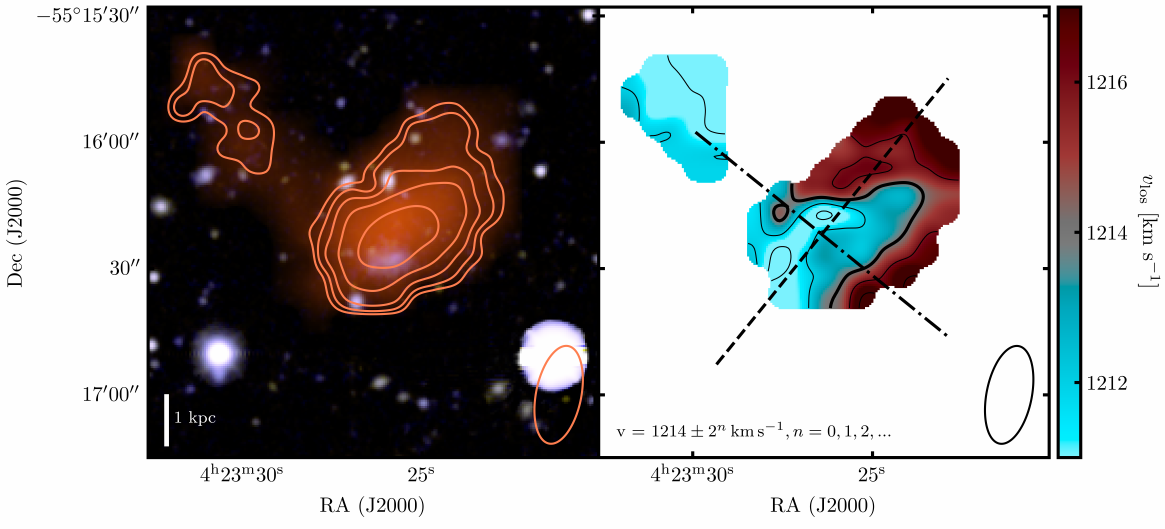}	
	\caption{{\em Left Panel}: DeCaLS optical image in the $g,r,z$ filters overlaid with the \HI\ contours from the $25$\arcsec$\times18$\arcsec datacube used for the analysis in this paper. Levels increase as $4.6\times10^{18}\times 2^n$~\cmsq\ (n=0,1,2,3), where the first contour marks the mean $S/N=3$ detection limit. The PSF is shown in the bottom-right corner. Note the \HI\ overlay in color with also the $32\times23$~\arcsec\ map. {\em Right Panel}: Velocity field of the \HI\ gas in \lsb. The systemic velocity ($v_{\rm sys} = 1214$~\kms) is marked by the thick black iso-velocity contour, the other contour levels are shown in the bottom-left corner. The dashed and dashed-dotted lines mark the directions of the major ($141^\circ$) and minor axes ($51^\circ$) of the \HI\ disk, respectively}.
	\label{fig:mom01}%
\end{figure*}

The left panel of Fig.~\ref{fig:mom01} shows the \HI\ and optical body of \lsb\ in detail. The first contour shows the $S/N=3$ flux-limit (before primary beam correction) at $25$\arcsec $\times 18$\arcsec, which corresponds to a column density of \nhi$=4.6\times10^{18}$~\cmsq.  Further details on the spatial and velocity extent of \lsb\ are visible in the channel maps (Fig.~\ref{fig:chanMaps}). The \HI\ gas is distributed in an irregular morphology that at first look resembles a disk with projected major axis $a=36$\arcsec\ ($D_\matHI = 2.8\pm 0.8$ kpc), and minor axis $b=22$\arcsec$=1.8\pm0.8$ kpc. The major axis is misaligned with respect to the optical body ($PA_\matHI=141^\circ$ and $PA_{\star}=190^\circ$). Along the minor axis, in the north-east, the \HI\ elongates in a low-column density cloud (\nhi$=3-9\times 10^{18}$~\cmsq). This cloud contributes to the blue-shifted tail in the \HI\ spectral profile (Fig.~\ref{fig:integratedSpectrum}), and is visible in the channel maps at velocities 1209~\kms and 1212~\kms. We highlight that also a low-mass \HI\ cloud without associated stars may be present (at the detection limit of our observations) $8\arcmin$ further north of \lsb\ (41-projected kpc) at coordinates RA $=04^{\rm h}$22$^{\rm m}$31$^{\rm s}$, Dec$= -55^{\rm d}$12$^{\rm m}$02$^{\rm s}$,  The integrated spectrum is shown in Fig.~\ref{fig:integratedSpectrumDark} and has full-width at zero intensity of $18$~\kms. The total estimated \HI\ mass is $3.65\times 10^5$~\msun, which is slightly higher than the $3\sigma$ detection limit over $16$~\kms\ of the observations.

The right panel of Fig.~\ref{fig:mom01} shows the velocity field of \lsb. The \HI\ gas is characterised by a gradient of velocities along the major axis. One interpretation is that regular rotating motions are ongoing in the gas. Nevertheless, some asymmetries are present in the velocity field with a deep V-shape penetrating the minor axis in the south-west. Independent of this, the morphology of the \HI\ shown in the left panel of Fig.~\ref{fig:mom01} is suggestive of a disk-like distribution. Under this assumption  we assume we can then infer an inclination from the axis ratio ($b/a$), following, for example~\citet{sakai:2000}:

\begin{equation}
    i = \cos^{-1}\sqrt{\frac{(b/a)^2-q_0^2}{1-q_0^2}}
\end{equation}

\noindent where $q_0=0.13-0.2$7 is the intrinsic minor-to-major axis ratio of edge-on spirals. Given the irregularity of the \HI\ morphology, we determine the range of possible inclinations assuming the extreme values of the $q_0$ range and propagating the errors on the axis, which we estimate to be half the size of the beam minor axis, $9$\arcsec (0.8 kpc). We obtain an inclination of $i=50^\circ \pm 15^\circ$. 

\begin{figure}
	\centering
	\includegraphics[width=\columnwidth]{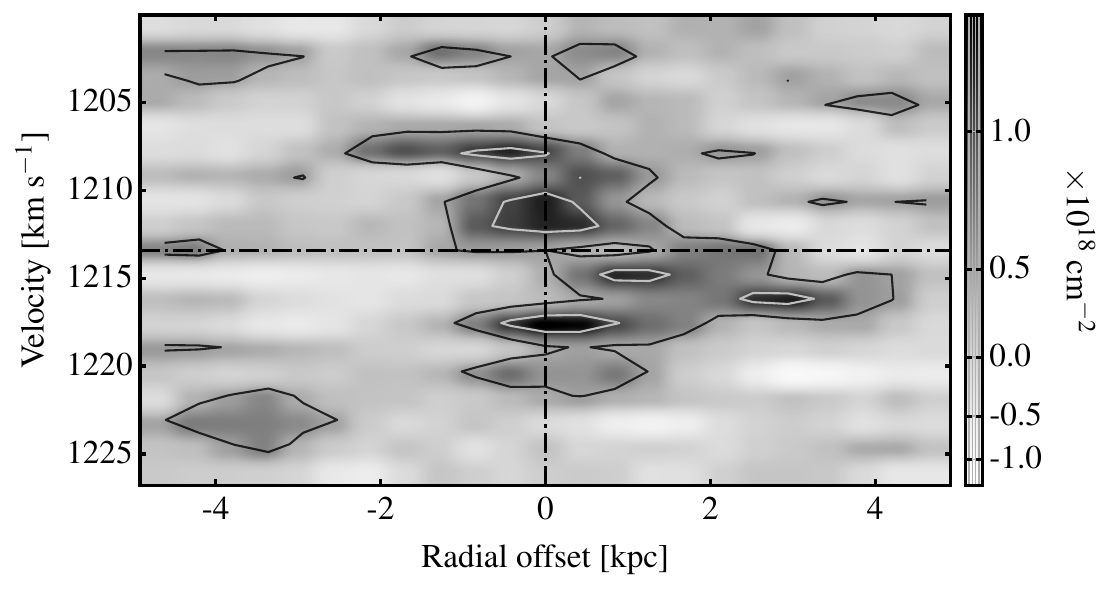}	
	\includegraphics[width=\columnwidth]{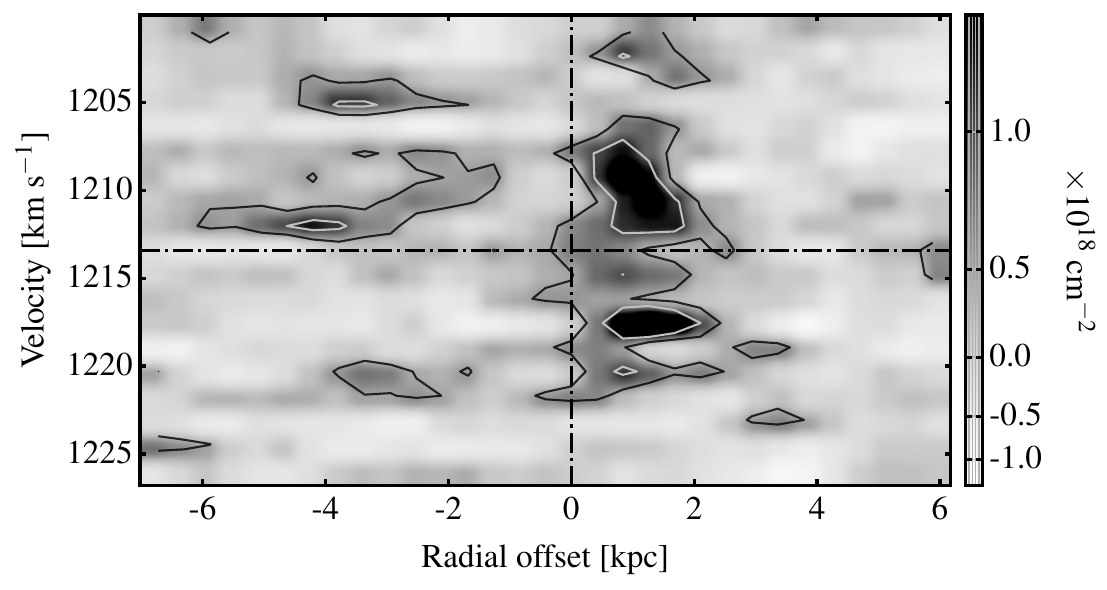}	
 \caption{{{\em Top Panel}: Position-velocity diagram taken along the major axis of the \HI\ disk ($PA=141^\circ$) with a slit of 20\arcsec. {Contour levels are at $5\times10^17$~\cmsq\ and $1\times10^18$~\cmsq}. {\em Bottom Panel}: Position-velocity diagram taken along the minor axis of the \HI\ disk ($PA=51^\circ$) with an analogous slit, equal to the resolution of the observations. Contours are as in the top panel.}}
	\label{fig:pvMaj}%
\end{figure}

To further investigate the kinematics of the \HI\ in \lsb, we draw the position-velocity (PV) diagram across the major (PA=$141^\circ$) and minor axis (PA=$51^\circ$) of the \HI\ disk (marked in the velocity field of Fig.~\ref{fig:mom01} with a dashed and dashed-dotted line, respectively). The PV-diagram across the major axis (Fig.~\ref{fig:pvMaj}, top panel) shows a steep gradient of velocities which disappears in the PV-diagram along the minor axis (bottom panel of the Figure), this may indicate that the \HI\ is supported by rotation. Along the velocity gradient, the \HI\ is detected only over three consecutive channels (of $1.4$~\kms\ each) indicating that the velocity dispersion is low ($\sigma=w_{50}/(2\sqrt{\ln{2}}) \lesssim 3\pm1.4$~\kms) as typically found for the cold \HI\ in rotating disks~\citep[$T<1000\,K$;][]{Young:2003,Warren:2012}. Under the assumption that rotation is (mostly) supporting the disk, a proxy for its rotational velocity can be inferred from the inclination-corrected velocity width~\citep{sakai:2000,begum:2008}: $v_{\rm rot} = 1/2\, w_{20}\sin(i)^{-1}(1+z)^{-1}$~\kms. From the smooth gradient of velocities across the major axis we measure $w_{20}=14$\kms\ and $v_{\rm rot}=9.0\pm3.5$~\kms. Because of the difficulty in estimating the rotational velocity we estimate an error of $\sim 40$\%. 

Figure~\ref{fig:pvMaj} shows that in the centre, a second peak is present at $v=1218$~\kms\ beyond the range of rotational velocities traced by the smooth gradient.  Without considering the beam-smearing due to the turnover velocities in the central resolution beam, the second peak causes the velocity dispersion to increase to $\sigma = 5.5$~\kms. This central velocity dispersion is typical of low-mass systems~\citep[\eg][]{Iorio:2017,Mancera2021b} and implies $v_{\rm rot}/\sigma_v \sim 2$. This suggests that the central region of the \HI\ disk has more turbulent kinematics, as also indicated by the shift of its centre with respect to the optical body (Fig.~\ref{fig:mom01}, left panel).

\begin{table}[tbh]
        \caption{\HI\ properties of \lsb}
        \centering
        \label{tab:properties}
        \begin{tabularx}{\columnwidth}{X l}  
                \hline\hline                   
                	Parameter   & Value  \\
                	\hline                                   
                	Right Ascension [J2000] &  $04^{\rm h}$23$^{\rm m}$26$^{\rm s}$ \\
					Declination [J2000] &  $-55^{\rm d}$16$^{\rm m}$29$^{\rm s}$\\
					$D$ [Mpc] & 17.69\\
                	\HI\ systemic velocity [radio,\kms] & 1214\\
                	Total flux density [mJy km s$^{-1}$] & $22.9\pm10\%$\\
                	M$_{\rm \HI}$ [\msun] & $1.68 \times 10^{6}$ \\
                	M$_{\rm bar}$  [\msun] & $4.68 \times 10^{6}$ \\
                    $v_{\rm rot}$ [\kms] & $9.0\pm3.5$ \\
                    $\sigma_{\matHI}$$^\dag$ [\kms] & $3\pm1.4$ \\
                	$PA$ (\HI\ disk position angle, $^\circ$) & $141 \pm 5$ \\
                	$i$ (\HI\ disk inclination, $^\circ$)  &  $50 \pm 15$\\
                	$a/2$ (\HI\ semi-major axis) [kpc] & $1.4\pm 0.8$ \\
                	$b/2$ (\HI\ semi-minor axis) [kpc] & $0.94\pm 0.8$ \\
                	f$_{gas}$ (gas fraction) & 0.48 \\
                	\hline                           
        \end{tabularx} 
         \tablefoot{$^\dag$: average value across the \HI\ disk, in the centre $\sigma=5.5\pm1$\kms.}
\end{table}

\subsection{Deep optical data and stellar properties of \lsb}
\label{sec:obsOpt}

\begin{figure}[tbh]
	\centering
	\includegraphics[width=\columnwidth]{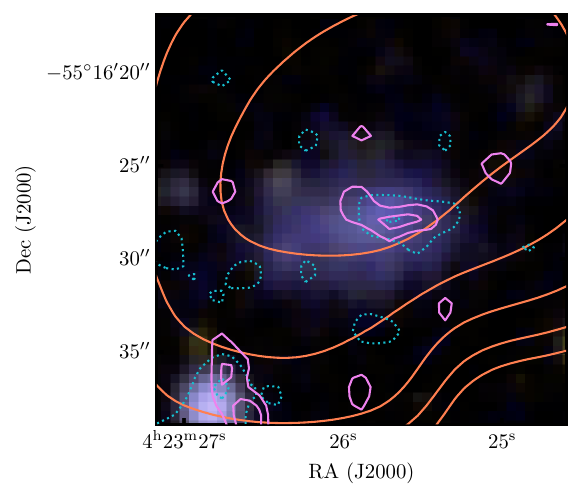}	
	\caption{VEGAS optical image in the $g,r$ filters overlaid with the \HI\ column density map. Orange contour levels are as the left panel of Fig.~\ref{fig:mom01}. In cyan and magenta contours we show the $S/N=3$ and $5$ contours from the NUV and FUV GALEX observations, respectively.}
	\label{fig:hi_uv}%
\end{figure}

Combining the deep \HI\ observations with optical photometric observations we can state that \lsb\ is without any reasonable doubt a gas rich low-surface brightness galaxy. To properly characterize its stellar properties we rely on the deepest available photometric observations of the Dorado group from  VEGAS~\citep[][]{Spavone:2017,Iodice:2021}. These observations reach sensitivities of the limiting surface brightness of $\mu_g = 28.9$ mag/arcsec$^2$, $\mu_r = 27.3$ mag/arcsec$^2$, in the $g$ and $r$ filters respectively, estimated for a point source at 5$\sigma$ over a circular area with FWHM$\sim$1\arcsec. These limits are approximately two magnitudes deeper than the DECaLS surface brightness limit~\citep[][]{Martinez-Delgado:2023}.
The VEGAS observations were acquired using the VST/OmegaCAM instrument (run ID: 0102.A-0669(A)), which has a total field of view (FoV) of $1^{\circ} \times 1^{\circ}$ and a spatial resolution of $0.214$ arcsec/pixel~\citep[]{kuijken:2011}. The images were reduced using the dedicated {\it AstroWISE}  pipeline \citep[for details see][]{McFarland:2013,venhola:2017} specifically developed for the reduction of VST/OmegaCAM observations. The $g$ and $r$ bands were analysed by following the same procedure described in, for example,~\citet{Ragusa:2021,Ragusa:2022b,Ragusa:2022a}, to which we refer for more details. In particular, the magnitudes are in the AB system, and are corrected for Galactic extinction using the coefficients from~\citet{Schlafly:2011}. The VEGAS $g,r$ image of \lsb\ is shown in Fig~\ref{fig:hi_uv}.

\begin{figure}[tbh]
	\centering
	\includegraphics[width=\columnwidth]{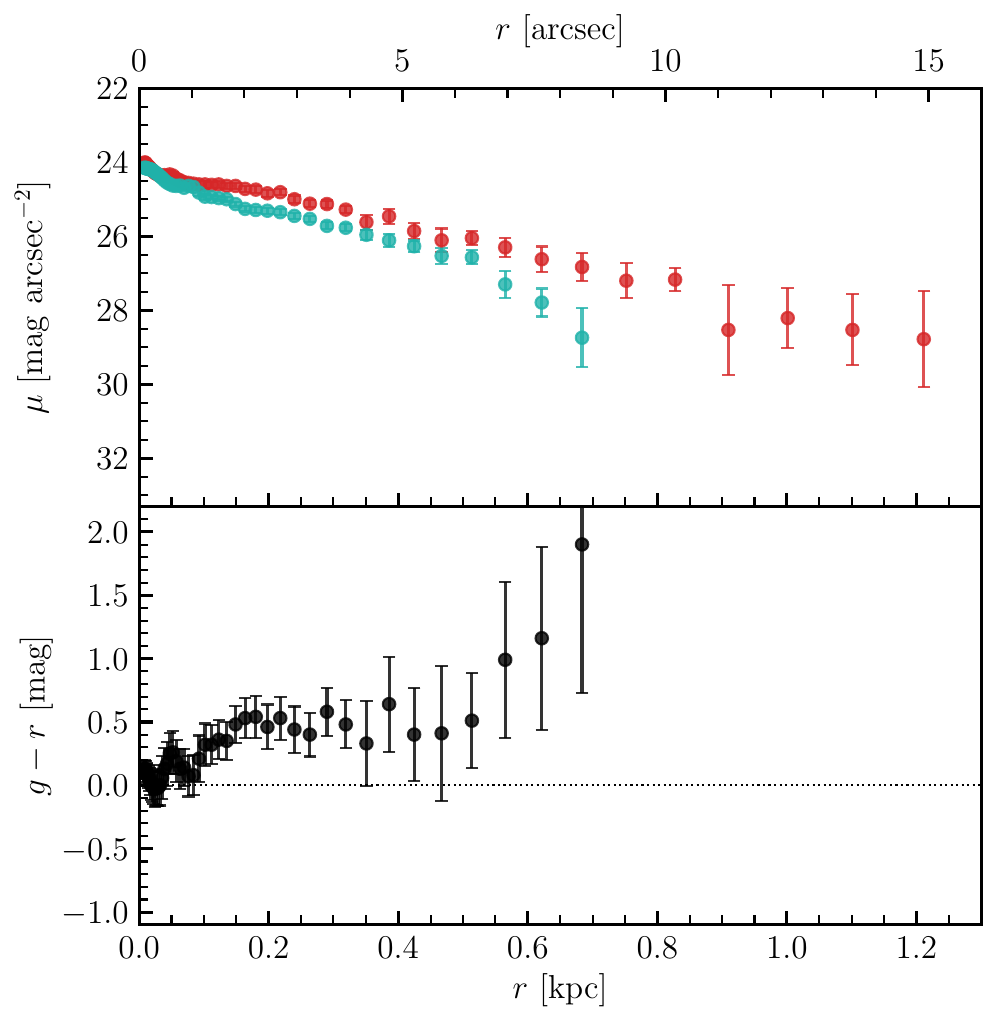}	
	\caption{{\em Top panel:} \lsb's $g$ and $r$ radial surface brightness profile (cyan and red, respectively) inferred from the VEGAS observations. {\em Bottom panel:} Radial $g-r$ colour profile measured out to $8$\arcsec\ from the centre ($\sim 670$ pc).}.
	\label{fig:vegasSB}%
\end{figure}

We infer the optical properties of \lsb\ with an isophote fitting procedure in the $g$ and $r$ bands. The surface brightness profiles are shown in the top panel of Fig.~\ref{fig:vegasSB}. The VEGAS observations confirm that the peak of the optical distribution of \lsb\ is shifted with respect to the \HI\ peak by $\sim 0.4$ kpc. We determine the stellar mass ($M_\star$) from $g$ and $r$ magnitudes ($M_r=-13.47$, $M_g=-13.04$) following the procedure illustrated in \cite{Into:2013}. Briefly, based on a mass-to-light ($M/L$) - color relationship, it is possible to derive an estimate of the average galaxy's $M/L$ by using its $g-r$ color. This, in turn, leads to an estimate, passing through the luminosity, of a stellar mass $M_\star\, (r) = 2.2\times10^6$~\msun\ and $M_\star\,(g) = 2.7\times10^6$~\msun, respectively. The stellar properties, along with the average $g-r$ colour, are listed in Table~\ref{tab:starProperties}.

\lsb\ has blue colours, similar to other nearby dwarf galaxies~\citep[\eg][]{honey:2018}, and small effective radius, $r_{\rm eff}=0.5$~kpc. Its radial colour profile (bottom panel of Fig.~\ref{fig:vegasSB}) shows a moderate reddening from the centre to the outskirts. Bluer colour in the centre suggest the presence of a younger stellar population or on-going star formation. 

To further investigate this possibility, we examine if ultraviolet (UV) emission is associated with \lsb\ in the {\em Galaxy Evolution Explorer Nearby Galaxy Survey}~\citep[][]{GildePaz:2007,Bianchi:2014}. Both in the far-UV ($\lambda\approx 154$~nm, FUV) and near-UV ($\lambda\approx231$~nm, NUV) observations emission is detected in the \HI\ disk of \lsb, see the magenta and cyan contours in Fig.~\ref{fig:hi_uv}. Given its low-signal-to-noise, the UV observations on their own would unlikely have been considered a detection. Since the peak is present both in the FUV and in the NUV and is coincident with the centre and the stellar body of the galaxy, we consider the detection significant.

The higher velocity dispersion of the central \HI\ gas may be due to the on-going or recent star formation suggested by the UV emission. The star formation rate (SFR) can be reliably estimated from the FUV emission~\citep[\eg][]{Leroy:2019}. We measure the UV-flux by centering a five-by-five pixel box on the peak, obtaining a total flux $F_{\rm FUV} = 0.031$ counts/sec. Correcting for extinction assuming a coefficient of $A_{\rm FUV} = 0.13$ from~\citet{Wall:2019} and given that the GALEX observation has an average background emission (as measured in a few areas nearby the source) of $F_{\rm backgroud}=0.00035$ counts/sec, the corrected flux is $F_{\rm FUV} = 0.025$ counts/sec. We determine the SFR from the correlation with the FUV, as inferred from the cross-matched sample of GALEX-SDSS galaxies~~\citep{Salim:2007,Leroy:2008}:

\begin{equation}
    {\rm SFR}^{\rm unobscured}_{\rm FUV} = 0.68 \times 10^{-28} L_\nu (\rm FUV) \quad {\rm M}_\odot \,\,\rm{yr}^{-1}
\end{equation}

\noindent where  $L_\nu(\rm FUV)$ is in \ergs Hz$^{-1}$.  For \lsb, $L_\nu(\rm FUV)=9.6\times10^{24}$~\ergs Hz$^{-1}$, which gives $\log({\rm SFR})_{\rm LSB-D}
 = -3.2$ [\msunyr]. \lsb's SFR value is a factor of only a few below the SFR of the lowest mass MHONGOOSE target galaxies (see Table 1 of~\citet{deBlok:2024}) and it is similar to the SFR in the low-mass end ($M_\star\lesssim 10^7$~\msun) of dwarf star-forming galaxies in the Local Universe~\citep[][]{McQuinn:2015a,McQuinn:2015b,Marasco:2023}.

\begin{table}[tbh]
        \caption{Stellar properties of \lsb}
        \centering
        \label{tab:starProperties}
        \begin{tabularx}{\columnwidth}{X l}  
                \hline\hline                   
                	Parameter   & Value  \\
                	\hline
					$M_r$ [mag] & $-13.04\pm0.25$ \\
					$M_g$ [mag] & $-13.47\pm0.11$ \\
                    $g-r$ [mag] &$0.46\pm0.21$ \\
					$M_\star (g)$ [\msun] & $2.2\times10^6$\\
					$M_\star (r)$ [\msun] & $2.7\times10^6$\\
                    $M/L (g)$  [\msun/\lsun]& 1.09 \\
                    $M/L (r)$  [\msun/\lsun]& 1.09 \\
                    $r_{eff}$ [kpc]& 0.5\\
                    $\mu_{eff}$ [mag/\arcsec] &26 \\
                    $L_\nu$ (FUV) [erg s$^{-1}$ Hz$^{-1}$] & $9.6\times10^{24}$\\
                    $\log {\rm(SFR_{\rm FUV})}^\dag$ [\msunyr] & -3.2\\
                	\hline                           
        \end{tabularx} 
\end{table}

\section{Discussion}
\label{sec:disc}

In this Section we discuss the possible evolution of \lsb\ and its \HI\ disk as it can be inferred from its location within the Dorado group (Sect.~\ref{sec:location}), by its gaseous and stellar content (Sect.~\ref{sec:dm}) and by its baryon-to-dark-matter content (Sect.~\ref{sec:bhr}). In Sect.~\ref{sec:evo}, we present possible scenarios on the formation and evolution of its \HI\ disk and on the enhancement of SF.

\subsection{\lsb\ in the Dorado group}
\label{sec:location}
\lsb\ lies $390$ projected-kpc to the north-east of the centre of Dorado (formed by the galaxy pair \mbox{NGC 1549/NGC 1553}), while the closest (in projection) massive galaxies \nfi\ and \mbox{NGC 1581} are $186$ kpc and $190$ kpc to the north-west and north-east, respectively. Since \lsb\ is at the edge of their projected virial radii it has likely not yet tidally interacted with them. Figure~\ref{fig:fullFieldCont} shows that \lsb\ is too far away from other more massive group members to be considered their satellite. The closest galaxy to \lsb\ is \mbox{MKT 042200-545440} (also associated for the first time to Dorado thanks to our MeerKAT observations) which is a late-type galaxy with similar \HI\ mass (see Table~\ref{tab:masses}).

\begin{figure}[tbh]
	\centering
	\includegraphics[width=\columnwidth]{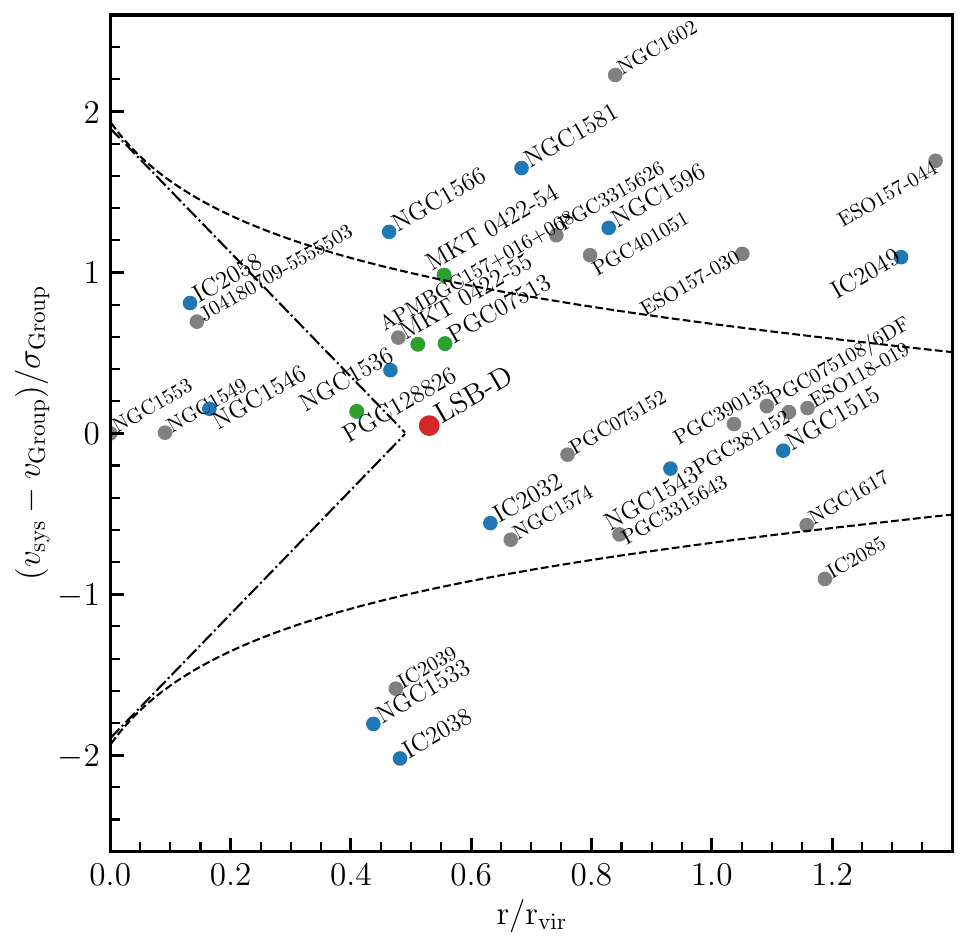}	
	\caption{Phase-space diagram of the Dorado group. \lsb\ is shown in red, while previously known \HI\ rich galaxies are shown in blue and new \HI\ detections in green. The group's centre (\mbox{NGC~1553}) is at the zero coordinates, the dashed-dotted lines mark the virialised region of the Dorado group. The caustic curves (dashed lines) are defined by the escape velocity of the group.}.
	\label{fig:phaseSpace}%
\end{figure}

The phase-space diagram of Dorado can provide further insights related to the location of \lsb\ and on its infall into the group. This diagram~\citep[see, for example,][]{Jaffe:2015} takes into account the  distance and systemic velocity of the group members as well as the virial mass and velocity dispersion of Dorado ($M_{\rm vir} = 3.5\times10^{10}$~\msun, $R_{\rm vir} = 528$ kpc, $\sigma_{\rm Dorado}=282$~\kms) to understand where the sources are located in phase space with respect to the BGG, if they are already within its escape velocity, if they recently assembled into the group or if they have long been virialised. Figure~\ref{fig:phaseSpace} shows the phase-space diagram of all Dorado members as catalogued in~\citet{rampazzo:2020,rampazzo:2021,rampazzo:2022} to which we add the sources newly associated with the group thanks to the \meer\ observations, \ie\ \lsb, \mbox{PGC 128826}, \mbox{PGC 075137}, \mbox{MKT 042200-545440}, \mbox{MKT 042230-550757}, \mbox{MKT 042326-551621}. The Figure shows the caustic curves defined by the escape velocity of the group (which is derived from its virial mass and radius) and the virialized area of the group (dashed-dotted lines). As illustrated in~\citet{Rhee:2017}, within this latter region it is more likely to find old members of the group between which interactions have already occurred, compared to recent infallers which are located between the virialized region and the caustic lines.

In the phase-space diagram, \lsb\ is within the escape velocity of the group in the region of recent infallers. The recent infallers are expected to be more gas-rich than similar systems in the virialized region, since the environmental effects of the group (\ie\ tidal interactions and/or ram pressure stripping) have not yet been strong enough or did not have enough time to deplete the sources of their gas.  

To further investigate if \lsb\ is a recent infaller of Dorado or if it had multiple orbits within the group, already, we estimate the probability distributions for orbital parameters and deprojected coordinates in group and clusters based on their phase space coordinates. In particular we follow the methodology illustrated in \citealt[][Sect.~3.2]{Oman:2021} (see also \citealp{Oman:2013}). In brief, from an N-body simulation catalogue we select within an overdensity similar to Dorado all orbits of satellites with projected coordinates, host mass and satellite mass compatible with \lsb. Further details on the selection constraints are given in Appendix~\ref{app:a2}. From the distribution of deprojected coordinates of the selection, we derive the probability for the deprojected coordinates of \lsb\ and for its orbit within the group. The median deprojected coordinates of \lsb\ (with $16-84$ percentile range) are $r_{\rm 3D}=506^{+481}_{-129}$ kpc, $v_{\rm 3D}=509^{+208}_{-242}$~\kms. These are shown in Fig.~\ref{fig:3dCoords} along with the distribution of the deprojected coordinates of all selected orbits. The probability for \lsb\ to be an interloper ($r_{\rm 3D}>2.5 R_{\rm vir}$) is $\lesssim 1$\%. The distribution of the pericentre of the selected orbits is bimodal with modes corresponding to either first or to second and later orbits. The probability for \lsb\ to be past apocentre is $\sim 50$\%, but if it was within the first orbit the probability for \lsb\ to be in the first infall before passing through the first pericentre is $\sim 27$\%, while the probability to be on the first orbit but past pericentre is $\sim 23$\%. In conclusion, the position in the phase space diagram and the predictions on its deprojected coordinates suggest that \lsb\ is a recent infaller of Dorado. The simulations suggest that in an overdensity similar to Dorado, galaxies with properties similar to \lsb\ with those projected coordinates are typically found on their first or at the beginning of the second orbit within the group.

\subsection{Gas and stellar content of \lsb\ and its star formation rate}
\label{sec:dm}

\begin{figure}[tbh]
	\centering
	\includegraphics[width=\columnwidth]{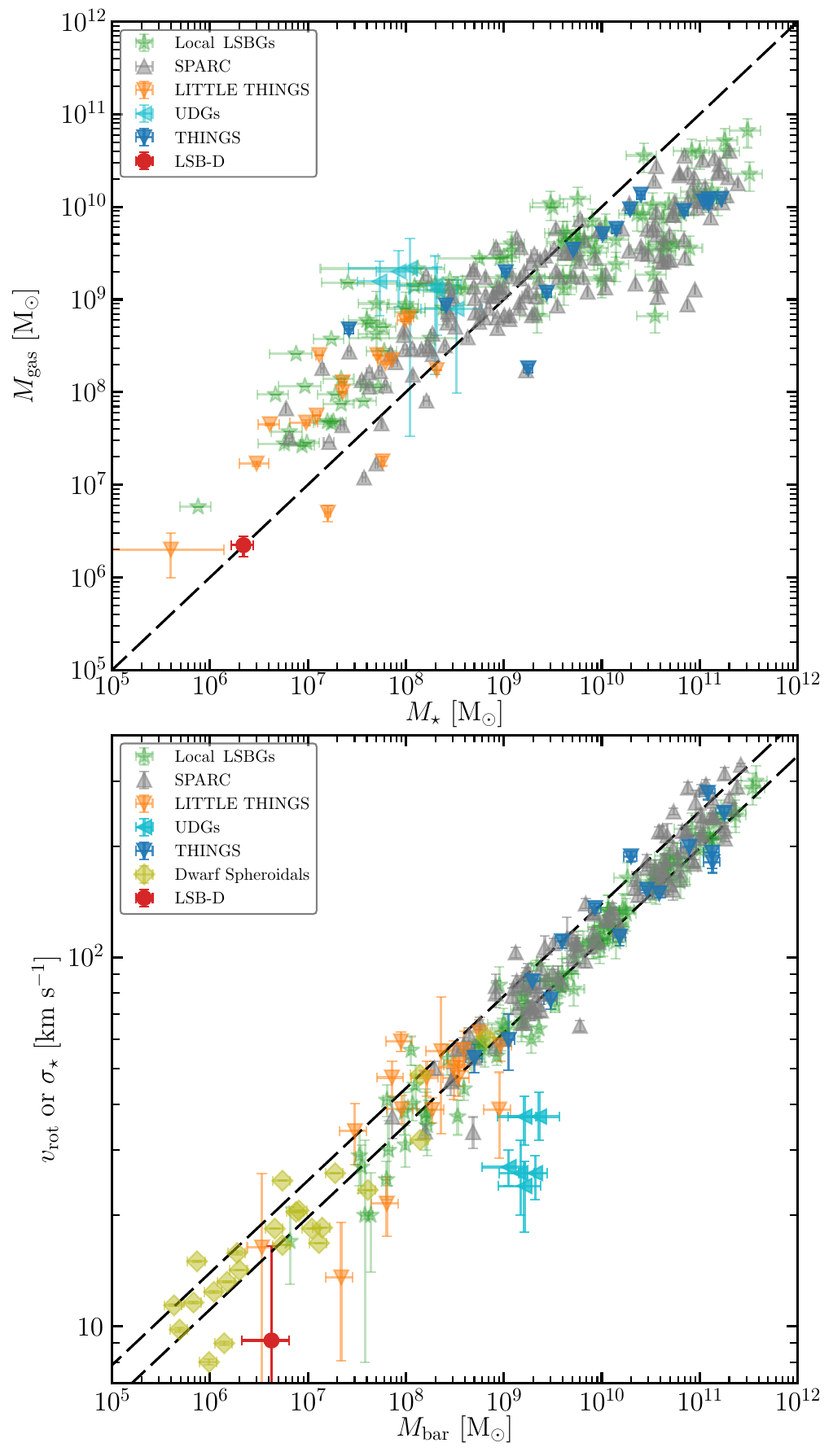}	
	\caption{{\em Top panel}: Gas mass versus stellar mass for a sample of galaxies in the Local Universe.  LITTLE-THINGS, SPARC and THINGS galaxies are marked by triangles (in orange, grey and blue, respectively). Local low surface brightness galaxies (LSBGs) from~\citet[][]{McGaugh:2011,McGaugh:2017,mcGaugh:2021} are in green, and UDGs in cyan. \lsb\ is marked by the red circle. The dashed line marks the $1:1$ linear relation. {\em Bottom panel}: The baryonic Tully–Fisher relation of Local Group dwarf and spiral galaxies~\citep[dashed line][]{mcGaugh:2021}. Symbols are as in the top panel. For the Local LSBGs sample, pressure supported dwarfs (plotted against their velocity dispersion, $\sigma_v$) are marked by light green squares, while green triangles mark rotationally-supported galaxies, plotted against their rotational velocity $v_{\rm rot}$.}.
	\label{fig:TF}%
\end{figure}

\lsb\ belongs to the low-mass end of the \HI-detected dwarf population. Here we compare its \HI\ properties with its stellar content and with those dwarf galaxies in the Local Universe for which we have reliable estimates. In particular, we select sources from LITTLE THINGS~\citep[][]{oh:2015,Iorio:2017}, SPARC~\citep[{\em Spitzer} Photometry and Accurate Rotation Curves,][]{Lelli:2016} and The \HI\ Nearby Galaxy Survey~\citep[THINGS,][]{Walter:2008} samples, local low surface brightness galaxies~\citep[from][]{McGaugh:2011,McGaugh:2017} and ultra-diffuse galaxies~\citep[UDGs, from][]{mancera2019,mancera2022_udg}, for which we have precise estimates of the gas and stellar masses, as well as rotational velocities. The top panel of Fig.~\ref{fig:TF} shows the stellar mass against the gas mass for the mentioned samples and for \lsb, which nicely fits on the one-to-one relation (shown by the dashed line). This Figure also highlights the low star and gas mass of \lsb\ of which only a few analogues in the Local Universe have been studied in detail. 

The baryonic Tully-Fisher (BTFR) relation links the baryonic mass of a galaxy with its rotational velocity~\citep[\eg][]{McGaugh:2000,Ponomareva2018,lelli:2019}. Whereas the relation is well established in the high mass regime~\citep[\eg][]{mcgaugh:2012,mcGaugh:2021,lelli:2022}, its nature in the low mass end is still a matter of heated debate as the scatter appears to increase towards lower masses~\citep{geha2006,begum:2008,mancera2019}. We investigate where \lsb\ is located on the BTFR to understand if dynamically its overall properties are typical of the ones of the low-mass low-surface brightness galaxies. 

In the bottom panel of Fig.~\ref{fig:TF} we show the BTFRs from~\citet{mcGaugh:2021}:

\begin{equation}
    v_0= 0.379\,\Bigg(\frac{M_{\rm bar}}{{\rm M}_\odot}\Bigg)^{1/4} \,\,{\rm km \,s}^{-1}
\end{equation}

\noindent The double dashed line in the Figure takes into account that there is a factor of two offset between the velocity estimated from rotation or pressure supported systems ($v_0$ and $\sigma$, details are given in~\citet{mcGaugh:2021}). In Fig.~\ref{fig:TF} we added the samples from LITTLE-THINGS, THINGS, and SPARC keeping their selection parameters and estimates of the rotational velocity and baryonic masses~\citep[there is a negligible discrepancy with the BTFR estimated with different methods for the $v_0$,][]{lelli:2019}. To this we also added the UDGs which are off the BTFR~\citep[][]{mancera2019}. \lsb\ (shown in red) seems to be slightly offset with respect to the relation, even though because of the large uncertainties (in the inclination and rotational velocity of the \HI\ disk) it can be consistent with the relation. We estimated \lsb's baryonic mass and rotational velocity as follows. Since we do not spatially resolve the source with enough resolution elements to properly characterize its rotation curve~\citep[\ie\ $\sim 5$ beams across the minor axis, see][]{Bosma:1978,Sancisi:2008}, we estimate a rough dynamical mass enclosed in the \lsb's \HI\ disk as $ M_{\rm dyn, rot}(\leq r_{\matHI}) \sim \frac{r_{\matHI}\,v_{\rm rot}^2}{G}\approx 2.6\pm1.0\times10^7$~\msun. Accounting for helium ($M_{\rm gas} = 1.33\cdot M_{\rm HI}=2.23\times10^6$~\msun), the baryonic mass is $M_{\rm bar}=M_{\rm gas}+M_\star=4.68\times10^6$~\msun\ and the total gas fraction is $f_{\rm gas}= M_{\rm gas}/(M_{\rm gas}+M_\star)=0.48$. \lsb\ is dark-matter dominated with its baryonic mass counting $5-10\%$ of the dynamical mass. This is compatible, even though slightly higher, with other low-surface brightness \HI\ rich galaxies in the Local Universe~\citep{giovanelli:2013,Iorio:2017}. Considering that \lsb\ may not be fully supported by rotation (since in the centre $v_{\rm rot}/\sigma_v \sim 2$), if we infer the dynamical mass from the maximum velocity dispersion through the virial theorem~\citep[as in other low-mass pressure supported systems, \eg][also shown in the Figure]{mcGaugh:2021}, we obtain a similar result because of the low velocities of the gas along the disk.   

The very low mass of \lsb, the normality of its gaseous and stellar properties and compatibility with the baryonic Tully-Fisher relation shows the universality of the dynamical properties of galaxies which defines their evolution throughout the entire spectrum of masses, down to the smallest low-surface brightness sources (albeit with larger scatter in this regime). 

 \begin{figure}[tbh]
	\centering
	\includegraphics[width=\columnwidth]{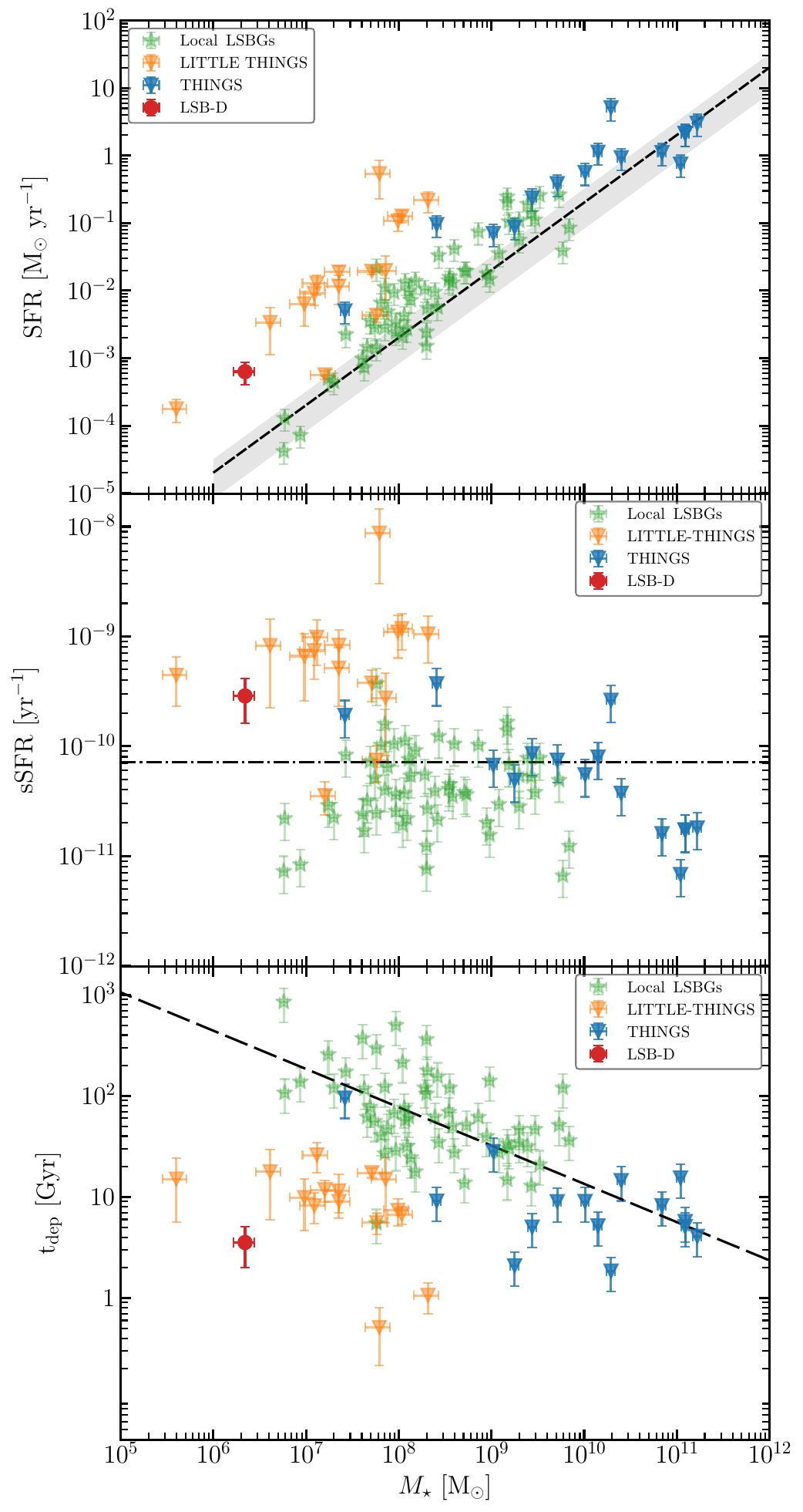}	
	\caption{{\em Top panel}: Star formation rate as a function of stellar mass for \lsb\ (marked by the red circle, and for a sample of galaxies in the Local Universe. LITTLE-THINGS~\citep[][]{oh:2015,Iorio:2017} and THINGS galaxies are marked by orange and blue triangles. Local low surface brightness galaxies (LSBGs) from~\citet[][]{McGaugh:2011,McGaugh:2017} are in green. The dashed line shows a linear relation fitted betwee the LSBGs and THINGS samples. {\em Central panel}: Specific star formation rate as a function of stellar mass for the same sample as in the top panel. The dashed-dotted line marks the inverse Hubble time, \ie\ galaxies on the relation reach their current stellar mass with their average SFR since the Big Bang. {\em Bottom panel}: Depletion time as a function of stellar mass for \lsb\ and the samples of galaxies in the Local Universe. The dashed line marks the linear fit relation for the Local LSBGs sample.}
	\label{fig:sfrRelations}%
\end{figure}

Overall, \lsb\ shows normal gas, stellar and dynamical properties compared to Local Group galaxies. This suggests that \lsb\ has undergone a `typical' evolution throughout cosmic time, and that past recent interactions or the environment of Dorado have not influenced its halo or mass content significantly. Nevertheless, the optical and UV properties of \lsb\ suggest, instead, that recently some event must have enhanced its star formation. Though gas rich by the standards of comparable low stellar mass galaxies (\ie\ dwarf spheroidals in Fig.~\ref{fig:TF}), \lsb\ has slightly lower gas masses than the Local LSBGs of not too much greater stellar mass. However, the top panel of Fig.~\ref{fig:sfrRelations} shows that \lsb\ is $\approx0.8$ dex above the $M_\star$-SFR relation defined by the Local LSBGs and THINGS galaxies. While it lies among the typically dwarf irregular LITTLE-THINGS galaxies\footnote{In this analysis we used the same samples of Fig.~\ref{fig:sfrRelations} for which measurements of the SFR and gas and stellar masses were available; SFR for THINGS are taken from~\citet{Leroy:2008}.}, some of the offset between \lsb, LITTLE-THINGS and the LSBGs and THINGS sample is due to the different tracers used to measure SF (FUV and H$\alpha$, respectively). For low mass galaxies with $\log({\rm SFR})\lesssim -2.5$~[\msunyr]\ this discrepancy can vary between a factor $2-10$~\citep[][]{Lee:2009}. However, in LITTLE-THINGS galaxies this discrepancy is at most a factor three, which is possibly due to a lack in sensitivity of the H$\alpha$ measurements~\citep[][]{Hunter:2010}. Part of the offset between LITTLE-THINGS galaxies and LSBGs may indicate that these dwarf irregulars have higher SFR than the Local LSBGs of comparable mass. Within this scenario, \lsb\ appears to be more similar to the SF dwarf irregulars than to the more quiescent LSBGs, which also tend to live in less dense environments.

When converting the SFR to specific star formation rate (${\rm SFR}/M_\star$) we see that while Local Universe galaxies lie around or below the inverse Hubble time relation, they would have reached their current stellar mass with their average sSFR since the Big Bang. \lsb\ lies above the relation suggesting that it would have doubled its stellar mass in a Hubble time with its current sSFR (Fig.~\ref{fig:sfrRelations}, central panel). Hence, for its stellar mass the current \lsb's SFR is too high to have always been on-going. Moreover, the depletion time of its gas reservoir ($t_{\rm dep} = M_{\rm gas}/{\rm  SFR}$) is much smaller than a Hubble time (Fig.~\ref{fig:sfrRelations}, bottom panel) indicating that if \lsb\ had formed stars at its current rate it would have already consumed all its gas.

\subsection{The baryon-to-dark-matter content of \lsb}
\label{sec:bhr}

\begin{figure}[tbh]
	\centering
	\includegraphics[width=\columnwidth]{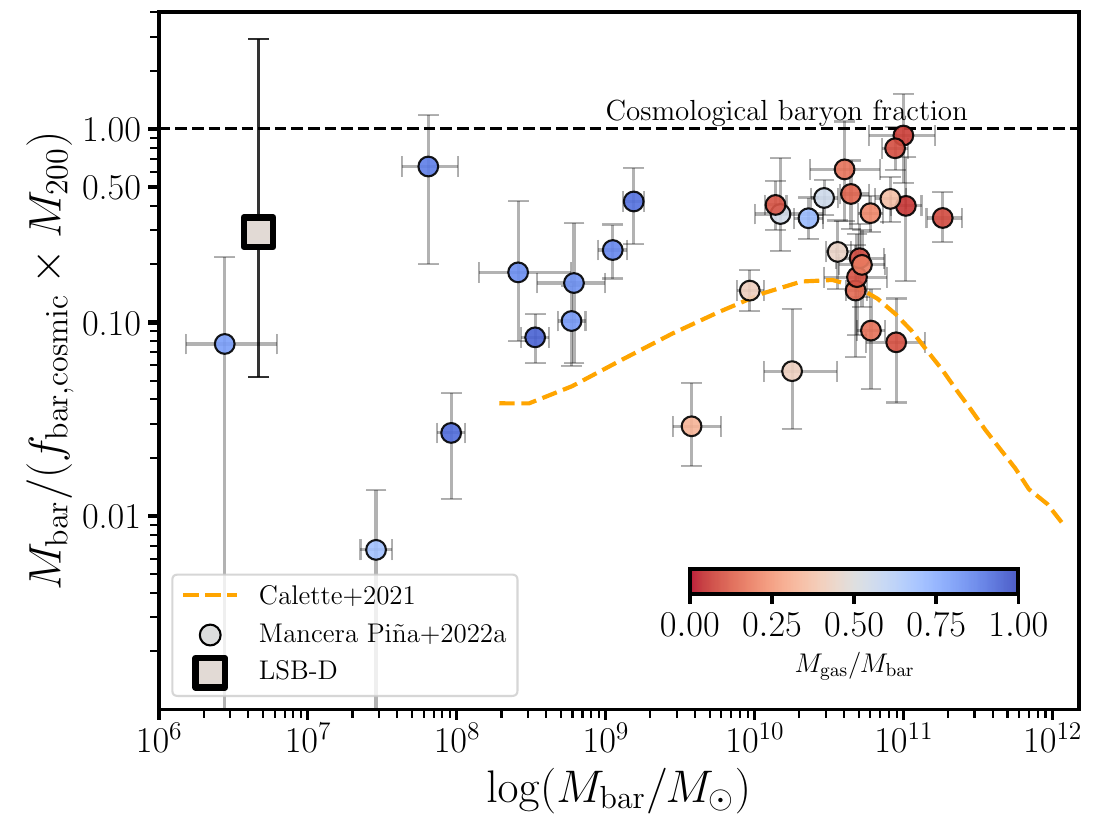}	
	\caption{$M_{\rm bar}/M_{200}$ ratio normalized to the cosmological baryon fraction ($f_{\rm bar, cosmic}$). \lsb\ is compared against a set of nearby disk galaxies \citep{Mancera:2022} and the semi-empirical relation from~\cite{calette:2021}, which uses abundance matching to link $M_\star$ with $M_{200}$ and empirical correlations for $M_{\rm star}/M_{\rm gas}$. Galaxies are colour coded according to their gas-to-baryon content ($M_{\rm gas}/M_{\rm bar}$).}
	\label{fig:bmhr}
\end{figure}

The baryon-to-halo mass relation of galaxies (BHMR) is a powerful tool to study the efficiency with which galaxies turn their available baryons into stars and cold gas. In Fig.~\ref{fig:bmhr}, we place \lsb\ on the BHMR and we compare it with expectations based on abundance-matching \citep{calette:2021} and against observations of a sample of dwarf and massive nearby disk galaxies spanning five orders of magnitude in baryonic mass from \cite{Mancera:2022}. For the latter galaxies the baryon fraction\footnote{$f_{\rm bar}$, the ratio between their baryonic ($M_{\rm bar}$) to halo ($M_{200}$) mass normalized by the cosmological baryon fraction $f_{\rm bar,cosmic}=0.16$.} is estimated via accurate mass modelling from their rotation curve decomposition~\citep[see][for details]{Mancera:2022}.

For \lsb\ we do not have a rotation curve, so we approximate its approximate location in Fig.~\ref{fig:bmhr} as follows. We estimate the circular speed of the galaxy as $v_{\rm circ} = v_{\rm rot}$. This means that we ignore corrections for line broadening due to turbulence as well as asymmetric drift (which would decrease and increase our $v_{\rm circ}$, respectively, e.g. \citealt{verheijen:2001,Iorio:2017}), which we cannot properly characterize because of the spatial resolution of our observations and considering that both corrections counteract each other. To estimate the uncertainty of $v_{\rm circ}$ we consider the error propagation in  $v_{\rm rot}$ (taking into account our error in the measurement in vrot and in the inclination uncertainty), and and we add (in quadrature) an additional term of 3~\kms\ associated with the asymmetric drift correction based on the typical value of such correction in low-mass systems \citep[\eg][]{lelli2014,Iorio:2017,Mancera2021b}. We compare our value of $v_{\rm circ}$ against a set of circular speed profiles of Navarro-Frenk-White~\citep[NFW,][]{Navarro:1996} haloes of different halo masses and with concentration parameters following the concentration-mass relation from N-body cosmological simulations \citep{dutton2014}. Given that this relation shows a tail towards low concentrations at low masses  \citep{mancera2022_udg,kong2022}, we also consider NFW profiles with concentration parameters $2\sigma$ below the nominal expectation. We find that our estimate of $v_{\rm circ}$ can be matched with the circular speed of a halo of $\sim 10^8~M_\odot$. Using this value, we add \lsb\ to Fig.~\ref{fig:bmhr}, though it is worth noticing that the uncertainties in $v_{\rm circ}$ allow for haloes in a broader mass range ($7 \lesssim \log(M_{200}/{\rm M}_{\odot}) \lesssim 8.75$). \lsb\ has baryon-to-halo mass properties similar to dwarf galaxy \mbox{DDO210}~\citep[\eg\ LITTLE THINGS,][]{Hunter:2012}.

Abundance matching techniques suggest that the BHMR relation steeply and monotonically increases as a function of $M_\star$ up to $M_{\rm halo}\sim 10^{12}$~\msun~\citep[\eg ][]{Moster:2013}, but the behaviour at low-masses is only extrapolated from the relation in the $10^9$-$10^{11}$~\msun\ range. The properties of \lsb, along with DDO210, suggest the relation may be flatter, and that low-surface brightness galaxies may retain their gas within the dark-matter halo while assembling their stellar content. This agrees with the BHMR relation obtained by \cite{calette:2021} who used both abundance matching to link $M_\star$ with $M_{200}$ and empirical correlations to link $M_\star$ to $M_{\rm gas}$, or the relation by \cite{posti:2019} based on \HI\ rotation curve measurements. Alternatively, \lsb\ and DDO210 may represent the part of the low-mass population with relatively large baryonic masses for their halo mass, while most of the sources with low baryonic masses are undetected.

\lsb\ and \mbox{DDO210} are two of the few known sources of low-stellar mass for which the stellar-to-mass halo content can be estimated accurately, hence they provide an observational constraint to the low-mass end ($\lesssim 5\times 10^7$~\msun) of the stellar-to-mass halo relation. Overall, our analysis on \lsb\ adds to the emerging picture of a large scatter in the baryon-content of low-mass galaxies due to feedback from star formation, which modern hydrodynamical simulations should aim to reproduce~\citep[][]{mcgaugh:2012}.

Even though the gravitational potential and the baryon content of \lsb\ and \mbox{DDO210} are similar, their \HI\ properties have some striking differences. First, the minimum \HI\ column density detected in \mbox{DDO210} is $6\times10^{19}$~\cmsq~\citep[which corresponds to a S/N=3 in the moment-zero map; ][]{Iorio:2017} and its \HI\ surface brightness profile peaks in the centre at $6\times10^{20}$~\cmsq. \lsb\ has \HI\ column densities one order of magnitude lower, peaking at $3.3\times10^{19}$~\cmsq, and it would have been at the detection limit of DDO210's VLA observations~\citep[\nhi($3\sigma,4$~\kms)$=1.3\times10^{19}$~\cmsq;][]{Hunter:2012}. Second, \lsb's \HI\ disk is much more extended ($r_{\matHI,\,\rm LSB-D}=1.4$~kpc), than the one of DDO210 ($r_{\matHI,\,\rm DD0210}=0.3$ kpc). \lsb\ and DDO210 likely had a different formation history, which one disk being much more diffuse than the other. \lsb\ appears closer in projection to its massive neighbours than DDO210. In the Local Group, DDO210 and other dwarf galaxies with \mhi$\gtrsim10^6$~\msun\ are located beyond $300$ kpc from the Milky Way or Andromeda. Within $300$~kpc dwarf satellites have low amounts of \HI\ (\mhi$\lesssim 10^6$~\msun) below the detection limit of currently available observations~\citep[\eg][]{Grcevich:2009,Spekkens:2014}, or no \HI\ at all. This also suggest that \lsb's interaction with the environment of Dorado, either began recently or has not been sufficiently disruptive to deplete the galaxy of \HI.

Different scenarios may explain the absence of galaxies similar to \lsb\ close to the Milky Way or Andromeda. Those low-surface brightness galaxies may not yet have entered the group (such as DDO210), or they could have already lost their \HI\ because of stronger interactions, or they could exist but because of their low-column density gas ($\lesssim 10^{19}$~\cmsq) they have not yet been detected. Dorado is not the only environment where we find several gas-rich low-mass satellites. Until wide-field \HI\ observations of the LG with comparable sensitivities and resolutions to MHONGOOSE will become available, it will be unclear if the evolution of Milky Way's satellites is representative of satellites in other groups, as suggested by the Satellites Around Galactic Analogs (SAGA) Survey~\citep[][]{geha:2017}, and if they provide a good representation of dwarfs in the local universe~\citep[][]{Weisz:2011}. Even though the \HI\ of the MW may interfere with the detection of close \HI\ rich satellites which could have low radial velocities and not have yet accreted into the galaxy~\citep[][]{Garrison-Kimmel:2014}.

\subsection{On the enhancement of star formation in \lsb}
\label{sec:evo}

\lsb\ is one of the few galaxies in the $10^6$~\msun\ stellar and gas mass range known beyond the Local Group. Its overall baryon and halo properties are typical of the dwarf population, nevertheless it presents two main peculiarities: the \HI\ disk has column densities on average one order of magnitude lower than the \HI\ disks of Local LSBGs ($\sim 10^{18-19}$~\cmsq) and in its centre UV emission indicates recent star formation. Here, we discuss on the possible evolutionary scenarios that shape \lsb\ as we see it now. 

\lsb\ is located in the outskirts of the Dorado group and likely only recently entered in the Dorado group (see Sect.~\ref{sec:location}). Possibly the interaction with the group's environment is linked to the enhancement of the star formation. With the information available we cannot connect the timescales of the infall to the triggering of the SF, but we can make some considerations on the environmental processes that may have occurred. \lsb\ could be the remnant of a tidal interaction that has occurred between \mbox{NGC 1581} and  \mbox{NGC 1566}, but tidal dwarfs usually lack dark matter and the distances and especially relative velocities between \lsb\ and these galaxies may suggest this is not the most likely scenario. The tidal interaction nevertheless strongly unsettled the \HI\ disks of the \mbox{NGC 1581} and \mbox{NGC 1566}. Several \HI\ streams and clouds are found around both galaxies (Fig.~\ref{fig:fullFieldCont}). In between \mbox{NGC 1581} and \lsb\ (at 45 projected-kpc from the latter) an \HI\ cloud is visible at the detection limit of our observations (see Sect.~\ref{sec:obsHI} and Fig.~\ref{fig:integratedSpectrumDark}) with small relative velocity to \lsb\ ($\Delta v=78$~\kms). The location of \lsb\ on the BHMR  (Sect.~\ref{sec:bhr}) suggests that the galaxy has a high fraction of baryons in its halo, $\sim 50\%$ of which are in the gas phase. It is possible that while entering the group, \lsb\ caught in its halo part of the \HI\ clouds remnant of the interaction. This \HI\ is now condensing onto the irregular disk we see now. This would explain why the disk may not fully rotating, the higher \HI\ line-width in the centre and the extended tail in the NW may be traces of this condensation. While condensing, part of the \HI\ in the centre may have triggered a new episode of SF. Nevertheless, very little is known about the evolution of these low-mass systems, numerical simulations have also shown that \HI\ disks of this mass can be retained by the galaxy's halo and re-settle in a rotating disk triggering a new episode of star-formation~\citep{Rey:2023}. In this picture, \lsb\ could also have retained its \HI\ within its disk over cosmic time and we are seeing a phase of re-settlement and condensation of its disk.  

Even though the origin of \lsb's \HI\ disk is unclear, different indications predict that it will be short-lived. Star formation feedback can disrupt an \HI\ disk similar to \lsb\ in a few hundred Myr~\citep{Rey:2023}, and the interaction with the intergalactic medium can also strip the gas off the galaxy. While most common in the dense environment of clusters, also in groups and the field ram pressure (RP) can be an efficient hydro-dynamical mechanism to strip galaxies~\citep[\eg][]{Poggianti:2017,Vulcani:2021}. RP is likely removing \HI\ from dwarf MW satellites and in the Sculptor group~\citep{Westmeier:2011,Westmeier:2015}, possibly triggering SF in the dwarf galaxies Pegasus, Wolf–Lundmark–Melotte and Holmberg II in the outskirts ($\sim 900$ kpc) of the LG and M81 groups~\citep[\eg][]{McConnachie:2007,Ianjamasimanana:2020,Yang:2022,Bureau:2002}. In the Fornax and  Virgo clusters, RP could have depleted of their gas dwarf galaxies with similar dynamical masses to \lsb~\citep[][]{Boselli:2008a,Venhola:2021,Kleiner:2023,Boselli:2023}. This would explain the abundance of dwarf gas poor ellipticals compared to other types of galaxies~\citep[\eg\ ][]{Boselli:2008b,Venhola:2021}. Interestingly, Dorado has tens of known low-surface brightness galaxies within 150 projected-kpc from the two central galaxies~\citep[][]{Carrasco:2001}. None are known to have gas, and the two that were in the \meer\ field of view had no detectable \HI. This may suggest that the hydro-dynamical interaction with the intra-group medium of Dorado has an effect on its low-mass members.

RP may already be shaping the \HI\ disk of \lsb. Ram pressure exerted by the IGM would explain the offset \HI\ disk with respect to the optical body, its elongated tail in the direction opposite of infall, and possibly the recent enhancement of the SF. RP is efficient when the pressure ($P_{\rm RP}$) exerted by the IGM on a galaxy moving at a relative velocity, $v$, overcomes the pressure imposed on the gas disk by a galaxy’s own gravitational potential~\citep[$P_{\rm grav}$,~\eg][]{gunn:1972}:

\begin{equation}
   P_{\rm RP} =\rho_{\rm IGM}v^2\gtrsim  2\pi G\Sigma_{\rm tot} \Sigma_{\rm gas} \simeq P_{\rm grav}
\end{equation}

\noindent where $\rho_{\rm IGM}$ is the IGM density, given by the product of the number of atoms in the IGM and the mean mass of a particle in the ionised medium, $\mu=0.75m_p$, $\Sigma_{\rm tot}=M_{\rm tot}/(\pi R_{\rm tot}^2)$ is the total surface density (stars plus gas) and $\Sigma_{\rm gas}$ is the ISM surface density of the disk. At the outermost edge of \lsb's disk ($r_\matHI=2.1$ kpc), the gas has column density of $2.8\times10^{18}$~\cmsq, which corresponds to $\Sigma_{\rm HI}=0.022$~\msun\ pc$^{-2}$. Correcting for helium abundance we obtain $\Sigma_{\rm gas}= 0.029$~\msun\ pc$^{-2}$. As illustrated in Section~\ref{sec:dm}, the total baryonic mass of \lsb\ is $M_{\rm tot} = M_{\rm bar} = 4.43\times10^6$~\msun. Hence, ram pressure would be effective on \lsb\ if it was greater than $P_{\rm RP} \gtrsim 2.9\times10^{-15}$ dyn cm$^{-2}$.

In Sect.~\ref{sec:location} we estimated the median deprojected velocity \lsb's velocity of \lsb\ based on a set of N-body simulation of satellites within a Dorado-like group, ($v=509^{208}_{-242}$~\kms). Assuming this velocity, for RP to take effect onto \lsb\ the IGM of Dorado would have to have densities $n_{\rm IGM} = \rho_{\rm IGM}\gtrsim 3\times10^{-7}$ cm$^{-3}$. In these low density environments ~\citep[similar to the outskirts of Virgo,][]{Steyrleithner:2020} RP has been observed to efficiently strip the gas off low \HI\ mass galaxies~\citep[\eg][]{Westmeier:2015,Ramatsoku:2019,Boselli:2023}, hence is possible that also in Dorado ram-pressure from the IGM is shaping the HI disk of \lsb.

Ram-pressure may enhance star-formation in the disks and in the stripped tails of galaxies~\citep[][]{Poggianti:2017,Vulcani:2018}, however beyond the Local Universe this mechanism has only been observed in galaxies of masses $M_\star~\gtrsim 10^9$~\msun~\citep[][]{Boselli:2008b,Fritz:2017}. In the outskirts of the Virgo cluster ($8\times10^{-5}\lesssim n_{\rm IGM}\lesssim 10^{-7}$~atoms~cm$^{-3}$), simulations have shown that SF can be caused by RP in galaxies with similar masses to \lsb\ ($M_{\rm bar}=10^6$~\msun, $M_{\rm DM}=10^8$~\msun) travelling through the medium with velocities as low as $200$~\kms~\citep[][]{Steyrleithner:2020}. Similarly, in the MW halo's outskirts RP of the same order of magnitudes ($\gtrsim 10^{-15}$ dyn cm$^{-2}$) may trigger SF in low-mass satellites~\citep[][]{Samuel:2023}. Possibly, the same phenomenon is on-going in \lsb.  If that were the case, it would be the first time that this phenomenon is observed in a source of this low dynamical mass in the outskirts of a low-density group, like Dorado. Stripping must have, nevertheless, only just recently begun, otherwise in a few tens Myr the galaxy would have likely lost its \HI, as is observed in the low-mass Fornax cluster~\citep[][]{Kleiner:2023}. This also highlights the short-lived fate of \lsb's \HI\ disk.  

\section{Conclusions and future prospects}
\label{sec:conc}
In this paper we presented new deep \meer\ 21-cm observations of the field of \nfi\ from MHONGOOSE which led to the discovery of the gas-rich low-surface brightness galaxy \lsb\ in the Dorado group. Combining the \HI\ observations with deep optical images from VEGAS we determined the \HI\ and stellar properties of one of the lowest mass \HI-rich galaxies thus far known outside the Local Group ($M_{\rm bar}=M_{\rm gas}+M_\star=4.43\times10^6$~\msun, $f_{\rm gas}=0.48$, Sect.~\ref{sec:obsHI}). The \HI\ disk has low-column densities, the gas is compressed in the southern edge and has an outer asymmetry in the opposite direction. Overall, the disk is likely supported by rotation. In the centre, where the \HI\ line-width is higher ($\sigma=5.5$~\kms, Sect.~\ref{sec:HI}), the stellar body has bluer colours and coincident UV emission indicates on-going star-formation (Sect.~\ref{sec:obsOpt}). The SFR of \lsb\ is similar to that of star-forming dwarf galaxies in the Local Group. 

The gas and stellar properties of \lsb\ allow us to draw a picture of its evolution within Dorado. In the past, \lsb\ likely had a first episode of star formation that shaped the stellar body as we see it now. During this episode and throughout the subsequent evolution through cosmic time, \lsb\ retained a high percentage of baryons and in particular \HI\ within the halo or this gas was accreted onto its halo, as indicated by its gas fraction and baryon-to-halo mass content (see Fig.~\ref{fig:bmhr}). Possibly, the \HI\ disk has not been consumed by further star formation because its column densities are too low for the gas to cool into dense molecular clouds and generate a new episode of star formation~\citep[][]{Kanekar:2011,maccagni:2017}. \lsb\ evolved mostly unperturbed until recent times, when it began its first infall into the Dorado group (Sect.~\ref{sec:location}). During its infall, \lsb\ did not have a major interaction with other nearby galaxies but likely the environment caused the formation of the \HI\ disk and the recent enhancement of SF. In Sect.~\ref{sec:evo}, we suggested that on the one hand, the orbit within the group may have caused the \HI\ in the halo to condense in the disk, and start a new epoch of star formation, or \lsb\ may have caught some sparse \HI\ clouds from Dorado's IGM which are now forming a disk, and stars. On the other hand, ram-pressure from the low-density IGM may also be stripping the \HI\ disk, while compressing it in the centre and possibly triggering the new episode of SF (Sect.~\ref{sec:evo}). In all scenarios, \lsb's \HI\ disk will likely be short-lived and survive a few hundreds of Myr only.

In the future, Integral Field Unit observations in the optical band may be able to shed new light on the star formation history of \lsb\ and provide information on its stellar kinematics. The high-resolution ($\lesssim 1''$) study of the stellar body of \lsb\ (and of its kinematics with respect to the \HI), as well as of its ionised gas and chemical composition, will enable us to obtain a precise measurement of its star formation rate, determine which phenomena affect its multi-phase ISM, and study its evolution within Dorado.

Given its very low mass, the discovery of \lsb\ beyond the Local Group, and the detailed study of its \HI\ and stellar properties demonstrates that the combination of MHONGOOSE observations with deep optical images opens a new parameter space of studies of low-surface brightness galaxies. In the observations presented in this paper, also four other sources with \HI\ mass lower than $2\times10^{7}$~\msun\ have been detected in the proximity of \nfi. MHONGOOSE is observing $\approx 24$ deg$^2$ with high sensitivity to low-density gas and low \HI\ masses, and is commensaly observing the same fields with the VST with high surface brightness sensitivities. This will enable the discovery and detailed analysis of several more low-mass gas rich galaxies in the field environment. The first data release (DR1, with 5 hours of observing time per target, $\sqrt{10}$ less sensitive than the observations shown in this paper) already discovered $5$ new sources~\citep[][]{deBlok:2024}, amongst the 46 additional \HI\ satellites detected around its targets. The full survey with ten times the observing time of MHONGOOSE-DR1 will certainly discover more objects like \lsb.

The studies of ultra low-mass \HI\ rich galaxies have so far been limited to the Local Group, which, alone, cannot provide a representative picture of the evolution of dwarf galaxies. The study of \lsb, and the many more that will be discovered in the future will make possible Local Group cosmological studies in different environments in the Nearby Universe. Extending the study done on \lsb\ to all newly discovered sources will allow us to populate the low-mass end of the BHMR and study the scatter (and its origin) in the BTF relation. Moreover, these studies enable a direct comparison with the kinematics, gas content and environmental effects of dwarf satellite galaxies present in large-volume hydrodynamical simulations~\citep[from, for example, TNG50, FIRE, APOSTLE in][respectively]{Rohr:2023,Samuel:2023,Jones:2023} which will quantify the main differences between the observed and simulated galaxy assembly processes, thus enabling us to provide observational constraints for a more detailed description of these phenomena in the next generation of simulations.

\begin{acknowledgements}
The authors thank the anonymous referee for the useful comments and suggestions.

We thank Dr. Teymoor Saifollahi for giving the suggestion to write this paper. 

This project has received funding from the European Research Council (ERC) under the European Union’s Horizon 2020 research and innovation programme (grant agreement no. 882793). 

PEMP acknowledges the support from the Dutch Research Council (NWO) through the Veni grant VI.Veni.222.364 
KAO acknowledges support by the Royal Society through a Dorothy Hodgkin Fellowship  (DHF\textbackslash R1\textbackslash 231105) and by STFC through grant ST/T000244/1. This work used the DiRAC@Durham facility managed by the Institute for Computational Cosmology on behalf of the STFC DiRAC HPC Facility (www.dirac.ac.uk). The equipment was funded by BEIS capital funding via STFC capital grants ST/K00042X/1, ST/P002293/1, ST/R002371/1 and ST/S002502/1, Durham University and STFC operations grant ST/R000832/1. DiRAC is part of the National e-Infastructure.
AB acknowledges support from the Centre National d’Etudes Spatial (CNES), France.
MK acknowledges support from the National Research Foundation of Korea (NRF) grant funded by the Korea government (Ministry of Science and ICT: MSIT) (No. NRF-RS-2023-00243222, NRF-RS-2022-00197685).
This paper makes use of MeerKAT data. The MeerKAT telescope is operated by the South African Radio Astronomy Observatory, which is a facility of the National Research Foundation, an agency of the Department of Science and Innovation. 

(Part of) the data published here have been reduced using the CARACal pipeline, partially supported by ERC Starting grant number 679627 “FORNAX”, MAECI Grant Number ZA18GR02, DST-NRF Grant Number 113121 as part of the ISARP Joint Research Scheme, and BMBF project 05A17PC2 for D-MeerKAT. Information about CARACal can be obtained online under the URL: https://caracal.readthedocs.io.
\end{acknowledgements}

\bibliographystyle{aa} 
\bibliography{LSBD-bib.bib} 
\begin{appendix}
\section{\HI\ properties of the sources detected in Dorado by MHONGOOSE}
\label{app:a1}
\begin{figure*}[tbh]
	\centering
	\includegraphics[width=\textwidth]{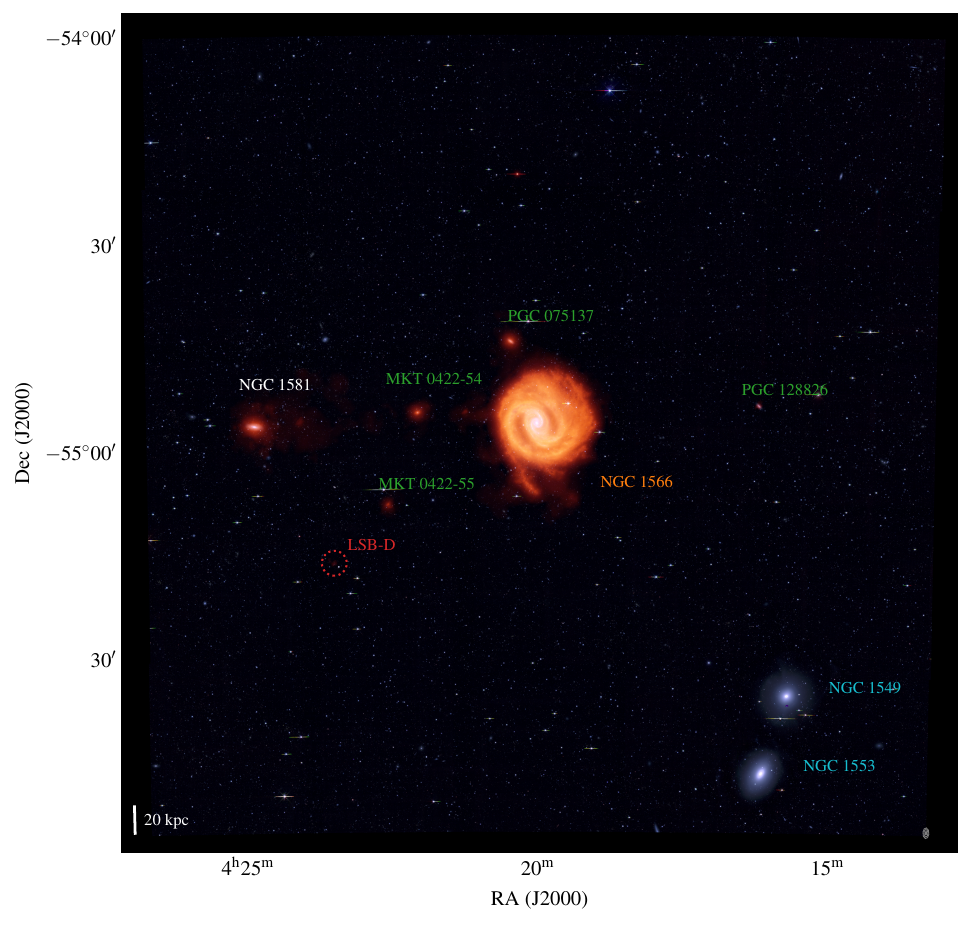}	
	\caption{Primary beam corrected flux-density \HI\ emission (in orange) detected by \meer, overlaid with the $2\times2$ degrees DECaLS optical image in the g,r,z filters. The \HI\ emission is an overlay of the flux-density maps extracted from the multi-resolution datacubes (see Sect.~\ref{sec:obs}). The PSFs of the multi-resolution maps varying from 90\arcsec\ to 7\arcsec\ are shown in white in the bottom-right corner. The target of the \meer\ observations, \nfi, is marked in orange, sources with previously known \HI\ are marked in white, while new \HI\ detections are marked in green. The doublet of early type galaxies in the centre of the Dorado group is marked in cyan. The newly discovered \HI\ low-surface brightness galaxy object of this paper is marked by the red circle. }
	\label{fig:fullField}%
\end{figure*}

As mentioned in Sect.~\ref{sec:obs}, we have identified six optimal weighting and tapering combinations that span the resolution range of MeerKAT ($7$\arcsec--$90$\arcsec) at optimal sensitivities with which we generate the \HI\ datacubes. The lowest resolution datacube ($94$\arcsec$\times92$\arcsec) is generated with Briggs {\em robust} $=1.0$ and taper to $90$\arcsec, hence is named {\tt r10\_t90}. With {\em robust} $=0.5$ and $60$\arcsec\ tapering we reach $65$\arcsec$\times64$\arcsec\ resolution ({\tt r05\_t60}). Higher resolution datacubes are generated with different robust weights ({\em robust} $=1.5,1.0,0.5,0.0$) and no tapering reaching resolutions of $32$\arcsec$\times23$\arcsec ({\tt r15}), $25$\arcsec$\times23$\arcsec\ ({\tt r10}), $12$\arcsec$\times10$\arcsec\ ({\tt r05}) and $7.6$\arcsec$\times7.2$\arcsec\ ({\tt r00}). The noise level matches the expected theoretical values, as illustrated in the survey paper~\citep[][]{deBlok:2024}. The $3\sigma$~\HI\ column density sensitivities in the centre of the cube are computed over $16$~\kms and vary between $5\times 10^{17}$~\cmsq\ at $90$\arcsec\ resolution and $5\times 10^{19}$~\cmsq\ at $7\arcsec$, for emission filling the beam uniformly. Figure~\ref{fig:fullField} shows the overlay of the \HI\ flux density maps derived from the datacubes at all resolutions. New \HI\ detections are marked in green.  Table~\ref{tab:cubes} summarizes the main properties of the datacubes. The datacube used to study \lsb's \HI\ disk is {\tt r10}, marked in bold in the Table. A summary of the \HI\ properties of the sources detected by the MHONGOOSE observations is given in Table~\ref{tab:masses}. 

\begin{table*}[tbh]
        \caption{Properties of the \mhon\ \HI\ datacubes of \nfi}
        \centering
        \label{tab:cubes}
        \begin{tabularx}{\textwidth}{X l c c c c } 
                \hline\hline                                                         
                Name & robust/taper	& Beam & Noise ($1\sigma)$ & $N_{\rm \HI}$ 3$\sigma$, 16~\kms & S/N=3    \\
       	             &           & [arcsec] & [\mJyb] & [$10^{18}$~\cmsq] & [$10^{18}$~\cmsq]   	                 \\
       	        \hline
               r10\_t90 &$r=1.0$ taper 90\arcsec & $94\times 92$ & $0.35$ & $0.55$ &$0.32$ \\
               r05\_t60  &$r=0.5$ taper 60\arcsec & $72\times 69$ & $0.27$ & $0.74$ &$0.61$ \\
               r15  &$r=1.5$ 					& $32\times 23$ &	$0.16$	& $2.9$ &$2.2$ \\ 
               {\bf r10}  &$r=1.0$ 				& 	$25\times 18$ &	$0.16$	& $4.9$ &$4.2$ \\ 
               r05  &$r=0.5$ 				& 	$12\times 10$ &	$0.18$	& $19$ &$19$ \\
               r00  &$r=0.0$ 				& 	$7.6\times 7.2$ &	$0.22$	& $51$ &$60$ \\
                \hline                           
        \end{tabularx}
\end{table*}

\begin{figure*}[tbh]
	\centering
	\includegraphics[width=\textwidth]{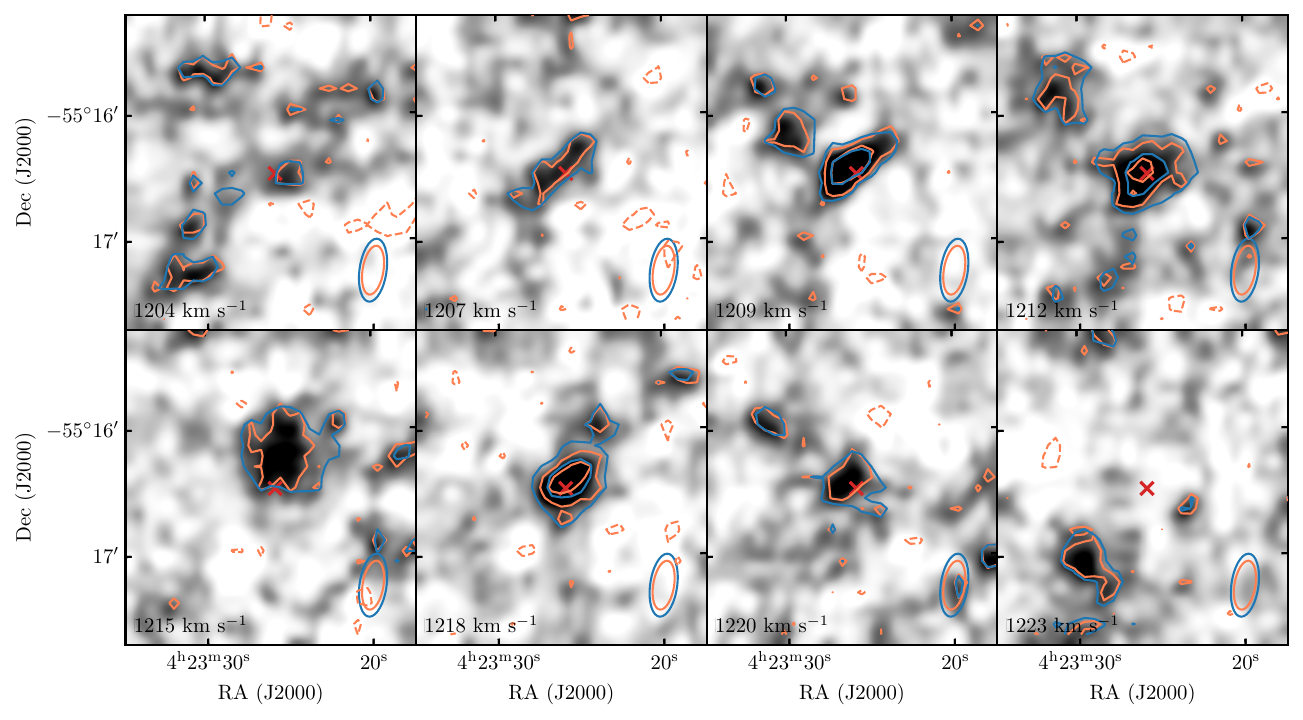}	
	\caption{Channel maps of the \HI\ emission of \lsb\ maps from the r10 and r15 datacubes (orange and blue contours, respectively). One every two channels is shown, contour levels increase as $2.5\sigma\times2^n$ ($n=0,1$). The red cross marks the centre of the optical body of the galaxy. The PSFs of the datacubes are shown in the bottom left corner.}
	\label{fig:chanMaps}%
\end{figure*}

 In Fig.~\ref{fig:chanMaps} we show the channel maps of the \HI\ emission of \lsb\ detected at $25$\arcsec$\times18$\arcsec and $32$\arcsec$\times23$\arcsec resolution. The centre of the optical body of the galaxy is marked by the red cross. One every two channel is shown, the  the \HI\ disk is visible between $1207$~\kms and $1220$~\kms, while the cloud in the north-east is visible at channels $1209$~\kms\ and $1212$\kms.

\begin{figure}[tbh]
	\centering
	\includegraphics[width=\columnwidth]{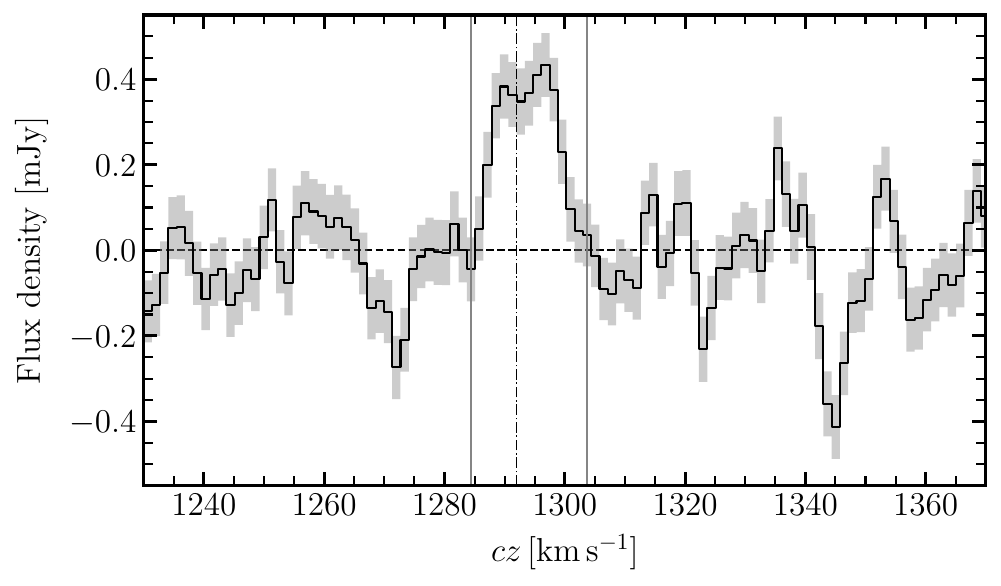}	
	\caption{Integrated spectrum the `dark' cloud detected at coordinates $04^{\rm h}22^{\rm m}31^{\rm s}$, $-55^{\rm d}12^{\rm m}02^{\rm s}$.  The vertical dashed-dotted line shows the central velocity of the \HI\ line ($v^{\rm radio}_{\rm sys}=1292$~\kms), while the grey vertical lines marks its full velocity range ($w_{\rm FWZI}=18$\kms).}
	\label{fig:integratedSpectrumDark}%
\end{figure}

Figure~\ref{fig:integratedSpectrumDark}, shows the integrated \HI\ profile of a `dark' cloud that may be present at the detection limit of our observations at coordinates: $04^{\rm h}22^{\rm m}31^{\rm s}$, $-55^{\rm d}12^{\rm m}02^{\rm s}$. At $25\times18$\arcsec, this emission line is unresolved, it is centred at \vsys$=1292$\kms\ and has full width at zero intensity of $w_{\rm FWZI}=18$~\kms. It is best detected after hanning-smoothing the datacube over three channels. The \HI\ integrated flux $S_{\rm 1.4 GHz}\Delta v=5.5$~\mJy~\kms. At the distance of Dorado, this corresponds to an 
\HI\ mass of \mhi$=3.6\times10^5$~\msun. This value is $20\%$ higher than the $S/N=3$ \HI\ mass detection limit for a source of $16$~\kms\ in those regions of the primary beam corrected datacube. If real this cloud may be the remnant of the past interaction between \nfi\ and \mbox{NGC 1581}. The projected distance between this \HI\ cloud and \lsb\ is $\sim 8.86$\arcmin\ ($\sim 45$ kpc) and their difference in line of sight velocities is $\Delta v=78$~\kms. Even though rare, similar `dark' low-column density  \HI\ clouds (\nhi$\lesssim 3\times 10^{19}$~\cmsq) have been detected as remnants of past interactions in groups~\citep[see, for example, ][]{Serra:2019,Maccagni:2023}. Some `dark' clouds also show indications of rotation~\citep[\eg][]{Xu:2023}, which could also occur in this newly detected cloud, as suggested by the double peaked profile.

\begin{table*}[tbh]
        \caption{\HI\ properties of the sources detected by MHONGOOSE in Dorado.}
        \centering
        \label{tab:masses}
        \begin{tabularx}{\textwidth}{X c c c c c} 
                \hline\hline                                                         
                Name	& Coordinates & $v_{\rm sys}$ & $F_{\rm int}$ & M$_{\rm \HI}^\dag$  & $w50$  \\
       	                & (J2000) & [\kms] &[mJy~\kms] & [\msun]    	           &[\kms]      \\
       	        \hline
                PGC 128826$^{*}$ &$04^{\rm h}16^{\rm m}17^{\rm s}-54^{\rm d}53^{\rm m}38^{\rm s}$& 1234 & 92.8 &$6.19\times 10^{6}$&84.0\\
                PGC 075137$^{*}$ & $04^{\rm h}20^{\rm m}26^{\rm s} -54^{\rm d} 44^{\rm m} 25^{\rm s}$& 1336 &251 &$1.82\times 10^{7}$&64.2\\
                \nfi & $04^{\rm h}20^{\rm m}42^{\rm s} -54^{\rm d}56^{\rm m}16^{\rm s}$ & 1496 &  $165\times 10^3$&$1.94\times 10^{10}$& 199\\
                MKT 042200-545440$^{*}$  & $04^{\rm h}22^{\rm m}00^{\rm s} -54^{\rm d}54^{\rm m}40^{\rm s}$ & 1439 &170 &$1.23\times10^{7}$& 29.2\\ 
                MKT 042230-550757$^{*}$  & $04^{\rm h}22^{\rm m}30^{\rm s} -55^{\rm d}07^{\rm m}57^{\rm s}$ & 1335 & 34.2&$2.49\times 10^{6}$& 14.0\\ 
                MKT 042231-551202$^{**}$ & $04^{\rm h}22^{\rm m}31^{\rm s} -55^{\rm d}12^{\rm m}02^{\rm s}$ & 1292 & 5.01&$3.65\times 10^{5}$& 10.0\\
                MKT 042326-551621$^{*}$, \lsb  & $04^{\rm h}23^{\rm m}26^{\rm s} -55^{\rm d}16^{\rm m}21^{\rm s}$ & 1214 & 22.9&$1.60\times 10^{6}$& 11.4\\
                \mbox{NGC 1581} & $04^{\rm h}24^{\rm m}44^{\rm s} -54^{\rm d}56^{\rm m}31^{\rm s}$ & 1609 &$1.38\times 10^{3}$ &$1.09\times 10^{8}$& 218\\ 
                \hline                           
        \end{tabularx}
        \tablefoot{($^{*}$) New \HI\ detections made by MHONGOOSE. ($^{**}$) \HI\ detection with $S/N=3$ without an optical counterpart. ($^{\dag}$) \HI\ masses are computed at the distance of \nfi, $D=17.69$ Mpc~\citep[][]{anand:2021}. }
\end{table*}

\section{Deprojected coordinates of \lsb\ derived from orbit catalogues}
\label{app:a2}

In Sect.~\ref{sec:disc} we provide the probability for \lsb\ to be a recent infaller of the Dorado group based on the probability distribution of its orbital parameters and deprojected coordinates of galaxies in groups and clusters of an N-body simulation. Here, were describe more in detail the methodology which follows~\citet[]{Oman:2021}. The input orbit catalogue is compiled from an N-body simulation of the same $(100\,\mathrm{Mpc})^3$ volume used in the EAGLE \citep{Schaye:2015} project (with a particle mass of $1.2\times 10^{7}\,\mathrm{M}_\odot$) run to a final scale factor of two. Dorado's virial mass and radius (see Table~\ref{tab:Dorado}) correspond to an overdensity $\Delta_\mathrm{vir}=M_\mathrm{vir} / (\rho_\mathrm{crit}\frac{4}{3}\pi r_\mathrm{vir}^3)=220$. Assuming a \citet{Navarro:1997} density profile and a concentration parameter $c_{200}=10$ we convert to the virial definition used in \citet{Oman:2021} and calculate the corresponding 3D velocity dispersion using their Eq.~1. The values used in this context are: $M_\mathrm{vir}=4.1\times 10^{13}\,\mathrm{M}_\odot$; $r_\mathrm{vir}=873\,\mathrm{kpc}$; $\sigma_\mathrm{3D}=571\,\mathrm{km}\,\mathrm{s}^{-1}$. Assuming \mbox{NGC 1553} as the group centre gives normalised offsets of $r/r_\mathrm{vir}=0.40$ and $(v_\mathrm{sys}-v_\mathrm{Group})/\sigma_\mathrm{3D}=0.02$. We select satellites from the orbit catalogue where: (i) the host mass is within $0.2\,\mathrm{dex}$ of $M_\mathrm{vir}$; (ii) the normalised radial offset is within $\pm0.1$ of that of \lsb; (iii) the absolute value of the normalised LoS velocity offset is less than $0.12$ (the absolute value reflects the degeneracy between the sign of the velocity offset and whether \lsb\ is in the foreground or background of the group); (iv) the satellite halo peak mass (maximum bound mass at any earlier time) is within $0.5\,\mathrm{dex}$ of $10^{8}\,\mathrm{M}_\odot$. The last element of the selection, (iv), is the most uncertain, but the \HI\ kinematics and stellar mass of \lsb\ suggest a total mass of about $10^{8}\,\mathrm{M}_\odot$ -- and the probability distributions obtained from the orbit library are only weakly sensitive to this choice.

 \begin{figure}[tbh]
	\centering
	\includegraphics[width=\columnwidth]{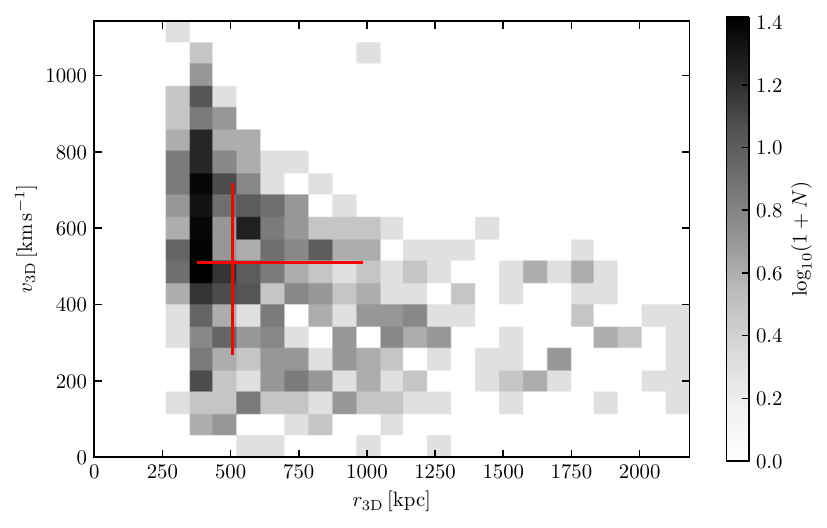}	
	\caption{Deprojected coordinates of LSB-D derived from orbit catalogues following the methodology of \citet{Oman:2021} as detailed in Sec.~\ref{sec:location}. The grayscale heatmap shows a histogram of the deprojected coordinates for simulated satellites with projected coordinates, host mass and satellite mass compatible with those of LSB-D. The red lines mark the median and $16^\mathrm{th}$-$84^\mathrm{th}$ percentiles of the marginalised probability distributions for the deprojected radius $r_\mathrm{3D}$ and current speed $v_\mathrm{3D}$.}
	\label{fig:3dCoords}%
\end{figure}

\end{appendix}

\end{document}